\documentclass[usenatbib, useAMS]{mn2e}
\usepackage{amsbsy}
\usepackage[dvips]{graphicx}
\usepackage{times}
\usepackage{amssymb}
\usepackage[authoryear]{natbib}

\makeatletter

\usepackage{times}

\newcommand{\aap}{A\&A}
\newcommand{\aaps}{A\&AS}
\newcommand{\apj}{ApJ}
\newcommand{\apjl}{\apj}
\newcommand{\pasp}{PASP}
\newcommand{\aj}{AJ}
\newcommand{\mnras}{MNRAS}
\newcommand{\apjs}{ApJS}

\newcommand{\pasj}{PASJ}
\newcommand{\araa}{ARA\&A}
\newcommand{\nar}{New Astron.~Rev.}

\newcommand{\kmps}{\mathrm{km~s^{-1}}}
\newcommand{\ion}[2]{#1$\,${\sc {#2}}}   
\newcommand{\Kelvin}{\mathrm{K}}
\newcommand{\Msun}{\mathrm{M_{\sun}}}
\newcommand{\Rsun}{\mathrm{R_{\sun}}}

\newcommand{\MsunPerYear}{\mathrm{M_{\sun}\,yr^{-1}}}

\makeatother

\begin{document}

\title[H and He line profile models of CTTSs]{Multidimensional models
  of hydrogen and helium emission line profiles for classical T\,Tauri
  Stars:  method, tests and examples} 

\author[R. Kurosawa et\,al.]{Ryuichi
  Kurosawa$^1$\thanks{E-mail:kurosawa@astro.cornell.edu},
  M.~M.~Romanova$^1$ and T.~J.~Harries$^2$ \\ $^1$Department of
  Astronomy, Cornell University, Space Sciences Building, Ithaca,
  NY 14853-6801, USA, \\ $^2$School of Physics, University of
  Exeter, Stocker Road, Exeter EX4~4QL, UK } 

\date{Dates to be inserted}

\maketitle

\begin{abstract}

We present multidimensional non-LTE radiative transfer models of
hydrogen and helium line profiles formed in the accretion flows and
the outflows near the star-disk interaction regions of classical T
Tauri stars (CTTSs). The statistical equilibrium calculations,
performed under the assumption of the Sobolev approximation using the
radiative transfer code {\sc torus}, has been improved to include
\ion{He}{i} and \ion{He}{ii} energy levels. This allows us to probe
the physical conditions of the inner wind of CTTSs by simultaneously
modelling the robust wind diagnostic line \ion{He}{i}~$\lambda10830$
and the accretion diagnostic lines such as Pa$\beta$, Br$\gamma$ and
\ion{He}{i}~$\lambda5876$.  The code has been tested in one and two
dimensional problems, and we have shown that the results
are in agreement with established codes.  We apply the
model to the complex flow geometries of CTTSs. Example model profiles
are computed using the combinations of (1)~magnetospheric accretion
and disc wind, and (2)~magnetospheric accretion and the stellar
wind. In both cases, the model produces line profiles which are qualitatively
similar to those found in observations.  Our models are consistent
with the scenario in which the narrow blueshifted absorption component
of \ion{He}{i}~$\lambda10830$ seen in observations is caused by a disc
wind, and the wider blueshifted absorption component (the P-Cygni
profile) is caused by a bipolar stellar wind. 
However, we do not have a strong constraint on the relative importance
of the wind and the magnetosphere for the `emission' component. 
Our preliminary calculations suggest that the temperature of the disc
wind and stellar winds cannot be much higher than $\sim$10,000~K, on
the basis of the strengths of hydrogen lines.  Similarly the
temperature of the magnetospheric accretion cannot be much higher than
$\sim$10,000~K. With these low temperatures, we find that the
photoionzation by high energy photons (e.g.~X-rays) is necessary to
produce \ion{He}{i}~$\lambda10830$ in emission and to produce the
blueshifted absorption components.

\end{abstract}


\section{Introduction}

\label{sec:intro}

Mass-loss processes in young stellar objects (YSOs), such as the
classical T Tauri stars (CTTSs), are a fundamental problem in the star
formation process.  The origin and the physical conditions of the wind
from YSOs are crucial to our understanding of the protostellar phase
and of the evolution of protostellar discs.  The mass-loss process is
closely related to the evolution of the stellar rotation since it is
one of the possible mechanisms through which YSOs may lose their angular
momentum (e.g.~\citealt{hartmann:1989};~\citealt{matt:2005};
\citeyear{Matt:2007b}; \citeyear{Matt:2008a}). The mass-loss due to
winds may solve the so-called `angular-momentum problem', in which a
large fraction of accreting CTTS have rather slow rotations
\citep{Herbst:2007} at $\sim10$~per~cent of break-up speed (e.g.,
~\citealt{Matt:2007}) despite of a rather large amount of the angular
momentum added to stars by accretion process.

Most likely, the outflows are formed in magnetohydrodynamics processes
in which a large-scale open magnetic field is anchored to a rotating
object; however, whether the object is the disk, central star or both
is still unclear (\citealt{Edwards:2006}; \citealt*{Ferreira:2006};
\citealt*{Kwan:2007}).  There are at least three possible
configurations of the wind formation: (1)~a disc wind in which one
assumes a sufficient magnetic field and ionization fraction to launch
magneto- centrifugal winds over a relatively wide range of disk radii,
from an inner truncation radius out to a several au
(e.g.~\citealt{ustyugova:1995}; \citealt{Romanova:1997}; 
\citealt{ouyed:1997}; \citealt{Ustyugova:1999};
\citealt{koenigl:2000}; \citealt{krasnopolsky:2003};
\citealt{Pudritz:2007}), (2)~an X-wind in which the wind launching region
is restricted to a narrow region near the inner edge of the accretion
disk around the corotation radius where the disk interfaces with the
stellar magnetosphere (\citealt{shu:1994}), and (3)~a stellar wind in
which outflows occurs along the open magnetic fields from the stellar
surface (e.g., \citealt*{hartmann:1982}; \citealt{Kwan:1988};
\citealt{Hirose:1997}; \citealt{Romanova:2005};
\citealt{Cranmer:2009}).  Interestingly, recent MHD simulations by
\citet{Romanova:2009} found yet another type outflow called `conical
winds' which are produced when the stellar dipole magnetic field is
squeezed by the disc into the X-wind type configuration. The resulting
outflow occurs in rather narrow conical shapes with their half opening
angles between $30^{\circ}$ to $40^{\circ}$. This conical wind is
similar to the X-wind mentioned above, but the former is driven by the
magnetic force and it does not required to have the magnetospheric
radius matched with the corotation radius of the system. 
More recent simulations performed with larger radial domains
have shown that the conical winds become strongly collimated by the
magnetic force at larger distances (\citealt*{Lii:2011};
\citealt*{koenigl:2011}).

In order to distinguish between the different theoretical models, and to understand
the origin of the angular momentum transfer, constraints
on the wind launching location are very important. Spectroscopic
studies of strong emission lines in CTTSs may help to solve this
problem. The observations by e.g.~\citet{Takami:2002}, \citet{Edwards:2003},
\citet{dupree:2005} and \citet{Edwards:2006} demonstrated a robustness
of optically thick \ion{He}{i}~$\lambda10830$ as a diagnostic of the
inner wind from the accreting stars. Unlike other strong emission lines
(e.g.~H$\alpha$ and H$\beta$) which also often show a sign of the
inner winds, \ion{He}{i}~$\lambda10830$ from CTTS shows P-Cygni type
profile (with deep blueshifted absorption below continuum level) which
resembles that of the hot stellar winds expanding in radial direction
(e.g., \citealt{Kwan:2007}). \citet{Edwards:2006} showed that about
70~per~cent of CTTS exhibit a blueshifted absorption (below continuum)
in contrast to H$\alpha$ which shows about only 10~per~cent of stars show
similar type of absorption component (e.g., \citealt*{reipurth:1996}).
\citet{Edwards:2007} and \citet{Kwan:2007} suggested that the
blueshifted absorption component in the \ion{He}{i}~$\lambda10830$
profiles is cased by a stellar wind in about 40~per~cent, and by a
disc wind in about 30~per~cent of the samples in \citet{Edwards:2006}. 
Recent local excitation calculations of \citet{Kwan:2011} in
combination with the spectroscopic observations demonstrated the
effectiveness of \ion{He}{i}~$\lambda10830$ for probing the density
and temperature of the inner wind.

In our previous radiative transfer models, (\citealt{harries:2000};
\citealt{kurosawa:2004}; \citealt*{symington:2005}; \citealt*{kurosawa:2006};
\citealt*{kurosawa:2008}), we have considered the statistical
equilibrium of hydrogen atoms only.  In light of the recent
recognition of \ion{He}{i}~$\lambda10830$ line as a useful wind
diagnostic tool (e.g.~\citealt{Edwards:2006};  \citealt{Kwan:2007};
\citealt{Kwan:2011}), we improve our model to include helium ions
(\ion{He}{i}, \ion{He}{ii} and  \ion{He}{iii}). This will allow as to 
investigate the physical conditions (e.g. geometry, temperature, 
density and kinematics) of the innermost part of the wind more
effectively, and will be a significant improvement to the wind study
of \citet{kurosawa:2006} who used only H$\alpha$. 
If we are to reliably determine mass loss rates and mass
accretion rates of CTTSs,  a more self-consistent modelling approach is
required. This includes a multiple line fitting of observations
including both hydrogen and helium to model the wind and
magnetospheric accretion flows simultaneously.

Our main goals in this paper are to present the methods and tests 
for our improved radiative transfer code, and to demonstrate the
capability of the code in the applications to the wind and
magnetospheric accretion problems of CTTSs. Detailed investigations
of the wind and accretions flows, e.g.~constraining flow geometry,
density, temperature and relative strength of the mass-loss to
mass-accretion rates, will be presented in a future paper as these
require a large parameter space survey.

In Section~\ref{sec:rt-code}, the model assumption, the method of
statistical equilibrium and the observed profile calculations are
presented.  The tests of the code in one and two dimensional (1D
and 2D respectively) geometries will be given in
Section~\ref{sec:results}. Simple kinematic models of a disc wind and a
stellar wind in combinations with a magnetospheric accretion model
will be presented in Section~\ref{sec:apps-to-ctts}. Example profiles of
hydrogen and helium are also given in the same section. Finally main
findings and conclusions are given in Section~\ref{sec:conclusions}.

\section{Radiative transfer code}

\label{sec:rt-code}

The radiative transfer code {\sc torus} (\citealt{harries:2000};
\citealt{kurosawa:2004}; \citealt{symington:2005}; \citealt{kurosawa:2006};
\citealt{kurosawa:2008}, but see also \citealt*{Acreman:2010}; \citealt{Rundle:2010})
is extended to include helium ions (\ion{He}{i}, \ion{He}{ii} and
\ion{He}{iii}). Our model uses the three-dimensional (3D) adaptive
mesh refinement (AMR) grid, and it allows us an accurate mapping of
an MHD simulation data on to the radiative transfer grid (without
a loss of resolution). The code also works in 2D (axisymmetric) and
1D (spherically symmetric) cases. The basic steps for computing the
line profiles are as follows: (1)~mapping of the density, velocity
and temperature, from either an MHD simulation output or an analytical
model, to the radiative transfer grid, (2)~the source function ($S_{\nu}$)
calculation and (3)~the observed flux/profile calculation. In the
second step, we adopt the method described by \citet{klein:1978}
(see also \citealt{rybicki:1978}; \citealt*{hartmann:1994}) in which
the Sobolev approximation method is used. The Sobolev approximation,
is valid when (1) a large velocity gradient is present in the gas
flow, and (2) the intrinsic line width is negligible compared to the
Doppler broadening of the line. In the following, we describe each
step of the line profile computation in more detail.

\subsection{Grid construction}

\label{sub:grid-construction}

Our radiative transfer code has been developed to handle length scales
in many order of magnitudes and multi-dimensional problems without
a particular symmetry (e.g., \citealt{kurosawa:2001}; \citealt{symington:2005};
\citealt{harries:2004}; \citealt{kurosawa:2004}). When the gradient
or dynamical range of physical quantities such as density is very
large, a logarithmic scale in the radial direction could be used to
increase an accuracy of optical depth calculations if
the system is spherically or axially symmetric. When a problem requires
an arbitrary geometry, there is no simple way to construct a logarithmic
scale. However, an efficient cubic or square grid (depending on the
problem is 3D or 2D) can be constructed by subdividing a cube/square
into smaller cubes/squares where the value of density or opacity is
larger than a given threshold value. We use the 8-way/4-way tree (octal/quad
tree) data structure to construct the grids and to store the physical
quantities required in the radiative transfer calculations. The data
structure used here is similar to a binary tree in which one node
splits into two child nodes, but ours splits into eight (for 3D) or
four (for 2D). 

\begin{figure}

\begin{center}
\includegraphics[clip,width=0.48\textwidth]{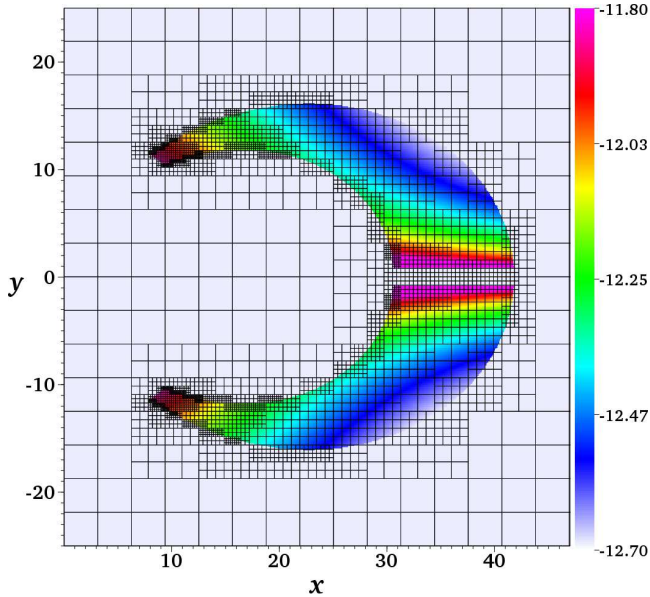}
\par\end{center}

\caption{An example of the 2D AMR grid constructed for the density
  from the magnetospheric accretion model described by
  \citet{hartmann:1994}. The grid structure is overlaid on the colour
  map of the input density.  The density map are logarithmically
  scaled and in cgs units. The axes are in $10^{10}$\,cm.   Although not
  shown here, the accreting star is located at ($x$, $y$) = ($0$,
  $0$), and its radius is $1.4\times 10^{11}$\,cm. }

\label{fig:grid-example}

\end{figure}

The algorithm used to construct the grid in this paper is very similar
to that used in \citet{kurosawa:2001}. Starting from a large cubic/square
cell (with size $D$) which contains a whole computational domain,
we first compute the average density of the cell by randomly choosing
100--1000 points in the cell and evaluating the density at each point.
The average density is then multiplied by the volume of the cell ($D^{3}$)
to find the total mass ($M_{\mathrm{cell}}$) in the cell. If the
mass is larger than a threshold mass ($M_{\mathrm{th}}$), which is
a user defined parameter, this cell is split into 8 or 4 subcells
with size $D/2$. If the mass of the cell is less than the threshold,
it will not be subdivided. The same procedure is applied to all the
subcells recursively, until all the cells contain a mass less than
the threshold ($M_{\mathrm{cell}}<M_{\mathrm{th}}$). Alternatively,
if the opacity of the field is known prior to the construction of
the grid (which is usually not the case), one can use the optical
depth across a cell ($\tau_{\mathrm{cell}}$) and use the condition,
$\tau_{\mathrm{cell}}<\tau_{\mathrm{th}}$ where $\tau_{\mathrm{th}}\approx0.1$,
as a condition for cell splitting. More complex algorithms could be
used to optimise the computation for a specific problem. Once the
data structure is built, the access/search for a given cell can be
done recursively, resulting a faster code. Fig.~\ref{fig:grid-example}
shows an example of the AMR grid constructed for the axisymmetric
(dipolar) magnetospheric accretion columns from \citet{hartmann:1994}. 


\subsection{Statistical equilibrium}

\label{sub:stateq}

Our statistical equilibrium computation essentially follows the method
described by \citet{klein:1978} (hereafter KC78; see also \citealt{rybicki:1978};
\citealt{hartmann:1994}) who uses the Sobolev approximation. A main
difference between the model of KC78 and ours is in the atomic model.
Our atomic model consists of 20 bound levels for \ion{H}{i} and the
continuum level for \ion{H}{ii}, 19 levels for \ion{He}{i} (up to
the principal quantum number $n=4$ level), 10 levels for \ion{He}{ii}
and the continuum level for \ion{He}{iii}. The atomic data used for
hydrogen are same as in \citet{harries:2000} and \citet{symington:2005},
and those for helium ions are adopted from \citet{hillier:1998}.
Here, we have a rather small set of ions and levels; however, the
method which are to be described next is not limited to these ions. 

Since we intend to apply our model to multi-dimensional problems (2D
and 3D), the main limiting factor of the size of the atomic model
is computer memory, and CPU time. For example, our typical 2D model
with CTTS accretion funnels and wind requires about $2\times10^{4}$
grid points ($\sim141^{2}$) and about $300$~Mb of RAM; hence adding
more complicated atoms and levels will be not feasible without a major
modification to the code. Note that our code is parallelised under
Message Passing Interface (MPI), but the domain decomposition of the
computational grid has not been implemented because the code is optimised
for speed. Adding more ions and modification of the code to handle
a grid with domain decomposition will be done in the future. 

In the original model of KC78, a simpler atomic model was used. They
assumed that a majority of the helium atoms were ionized, and excluded
\ion{He}{i} levels in their calculation. Their assumption is acceptable
since their objective is to model emission lines in the hot wind of
Of stars which have $T_{\mathrm{eff}}\gtrsim$30,000~K. The same
assumption can not be applied to flows with much lower temperatures
such as in the CTTS accretion flows and wind; hence, we have added
those extra levels. 

In the following, we summarise our computational steps. The following
equation of statistical equilibrium is applied to each levels of \ion{H}{i},
\ion{He}{i} and \ion{He}{ii}. The total rates for $j$-th level
can be written as:
\begin{equation}
  R_{j}^{L}-R_{j}^{U}+R_{j}^{R}-R_{j}^{I}=0,
  \label{eq:stateq}
\end{equation}
where $R_{j}^{L}$ is the net transition rate between $j$-th level
and the lower bound levels, and similarly $R_{j}^{U}$ is the net
transition rate between $j$-th level and the upper bound levels.
The last two terms, $R_{j}^{R}$ and $R_{j}^{I}$, are the net recombination
and the ionization rates, respectively. The rates in equation~(\ref{eq:stateq})
are:
\begin{eqnarray}
  R_{j}^{L} & = & \sum_{i<j}\left[n_{i}\left(B_{ij}\mathcal{J}_{ij}+n_{e}C_{ij}\right)\right.\nonumber \\
    &  & \left.-n_{j}\left(A_{ji}+B_{ji}\mathcal{J}_{ij}+n_{e}C_{ji}\right)\right]\,,\label{eq:rate-lower}\\
  R_{j}^{U} & = & \sum_{i>j}\Bigl[n_{i}\left(A_{ij}+B_{ij}\mathcal{J}_{ij}+n_{e}C_{ij}\right)\nonumber \\
    &  & -n_{j}\left(B_{ji}\mathcal{J}_{ij}+n_{e}C_{ji}\right)\Bigr]\,,\label{eq:rate-upper}\\
  R_{j}^{R} & = & n_{k}\left(\frac{n_{j}^{*}}{n_{k}^{*}}\right)\Bigl[4\pi\int_{\nu_{j}}^{\infty}\frac{a_{j}\left(\nu\right)}{h\nu}\left(\frac{2h\nu^{3}}{c^{2}}+J_{\nu}\right)\nonumber \\
    &  & \times\exp\left(-\frac{h\nu}{kT}\right)d\nu+n_{e}C_{jk}\Bigr]\,,\label{eq:rate-recombine}
\end{eqnarray}
and
\begin{equation}
  R_{j}^{I}=n_{j}\left(4\pi\int_{\nu_{j}}^{\infty}\frac{a_{j}\left(\nu\right)}{h\nu}J_{\nu}d\nu+n_{e}C_{jk}\right),
  \label{eq:rate-ionize}
\end{equation}
where $A_{ij}$ and $B_{ij}$ are the Einstein coefficients between
levels $i$ and $j$, and $C_{ij}$ is the collisional rate coefficient.
The first and the second subscripts in $A$, $B$ and $C$ correspond
to the initial and the final state of a transition, respectively.
The subscript $k$ denotes the continuum state. The LTE and non-LTE
populations of $i$-th level are written as $n_{i}^{*}$ and $n_{i}$,
respectively. While $a_{j}\left(\nu\right)$ is the photoionization
cross section of $i$-th level, $J_{\nu}$ is the mean (angle averaged)
intensity of the continuum radiation. The symbol $\mathcal{J}_{ij}$
refers to the angle and line profile averaged intensity of the line
transition between levels $i$ and $j$ (cf.~\citealt{mihalas:1978}). 
The electron number density and temperature are specified as $n_{e}$
and $T$. Other symbols have usual meaning. Using the Sobolev approximation
(the escape probability theory), $\mathcal{J}_{ij}$, for the transitions
with $i<j$, can be simplified as (see e.g.~\citealt{castor:1970}): 
\begin{equation}
  \mathcal{J}_{ij}=\left(1-\beta_{ij}\right)\frac{2h\nu_{ij}^{3}}{c^{2}}\left(\frac{g_{j}}{g_{i}}\frac{n_{j}}{n_{i}}-1\right)^{-1}+\beta_{c,ij}I_{c,ij},
  \label{eq:J-averaged}
\end{equation}
where $I_{c,ij}$ is the intensity of the continuum around the line
frequency $\nu_{ij}$. The symbols $g_{i}$ and $g_{j}$ denote the
degeneracy (statistical weight) of the energy levels. Here, we assume
that the `continuum emissivity and opacity of the gas are very small,
and the photospheric photons are unaffected by continuum processes.
In equation~(\ref{eq:J-averaged}), we neglect the contribution from the
non-local emission which are included in the models of
\citet{hartmann:1994} and \cite{muzerolle:2001}. The effect of this
omission is very small as we will see later in Section~\ref{sub:test-2D}. 
The escape probabilities $\beta_{ij}$ and $\beta_{c,ij}$ in equation~(\ref{eq:J-averaged})
are (c.f.~\citealt{castor:1970}; \citealt{rybicki:1978}): 
\begin{equation}
  \beta_{ij}=\frac{1}{4\pi}\oint\frac{1-e^{-\tau_{ij}}}{\tau_{ij}}d\Omega,
  \label{eq:beta-line}
\end{equation}
and 
\begin{equation}
  \beta_{c,ij}=\int_{\Omega_{*}}\frac{1-e^{-\tau_{ij}}}{\tau_{ij}}d\Omega 
  \label{eq:beta-cont}
\end{equation}
respectively. In equation~(\ref{eq:beta-cont}), the integration is performed
over the solid angle of stellar photosphere ($\Omega_{*}$) as
seen from the location where $\beta_{c,ij}$ is evaluated. The optical
depth $\tau_{ij}$ is defined as
\begin{equation}
  \tau_{ij}=c\frac{\chi_{ij}}{\nu_{ij}}\left|\frac{dv_{n}}{dl_{n}}\right|^{-1}.
  \label{eq:tau-sob}
\end{equation}
Here, $dv_{n}/dl_{n}$ is the derivative of the velocity field projected
in the direction $\hat{\boldsymbol{n}}$, i.e.~$v_{n}=\boldsymbol{v}\cdot\hat{\boldsymbol{n}}$,
and $dl_{n}$ is the line element in the direction of $\hat{\boldsymbol{n}}$.
The line opacity $\chi_{ij}$ is: 
\begin{equation}
  \chi_{ij}=\frac{\pi
    e^{2}f_{ij}n_{j}}{m_{e}c}\left(1-\frac{g_{j}n_{i}}{g_{i}n_{j}}\right), 
  \label{eq:chi-sob}
\end{equation}
where $f_{ij}$ is the oscillator strength.

Since we assume that the gas is optically thin to the photospheric
continuum radiation, the following simplifications are used for the
continuum intensity in equation~(\ref{eq:J-averaged}) and the mean
intensity in equations~(\ref{eq:rate-recombine}) and (\ref{eq:rate-ionize}).
We use the plane-parallel model atmosphere (the Eddington flux $H_{\nu}$)
of \citet{kurucz:1979}, then they become: 
\begin{equation}
  I_{c,ij}=4H_{\nu_{ij}}
  \label{eq:i-continuum}
\end{equation}
 and 
\begin{equation}
  J_{\nu}=4WH_{\nu},
  \label{eq:j-simple}
\end{equation}
where $W$ is the geometrical dilution factor $W=[1-\left(1-R_{*}^{2}/r^{2}\right)^{1/2}]/2$
(cf.~\citealt{mihalas:1978}). Alternatively, we can assume that
the photosphere is a blackbody (as in \citealt{hartmann:1994}; \citealt{muzerolle:2001}),
then we simply have $I_{c,ij}=B_{\nu}\left(T\right)$ and $J_{\nu}=WB_{\nu}\left(T\right)$
with an appropriate photospheric temperature $T$. 

The hydrogen particle number conservation gives
\begin{equation}
  \sum_{i=1}^{N_{\mathrm{HI}}}n_{i}\left(\mathrm{H\,
    I}\right)+n\left(\mathrm{H\, II}\right)=\frac{\rho\,
    X}{\left(X+4Y\right)m_{p}},
  \label{eq:hyd_num_cons}
\end{equation}
where $X$ and $Y$ are the abundances of hydrogen and helium (by
number), and $m_{p}$ is the proton mass. The symbol $N_{\mathrm{HI}}$
denotes the number of \ion{H}{i} levels (20). Similarly, the helium
particle number conservation gives
\begin{eqnarray}
  \sum_{i=1}^{N_{\mathrm{HeI}}}n_{i}\left(\mathrm{He\,
    I}\right)+\sum_{i=1}^{N_{\mathrm{HeII}}}n_{i}\left(\mathrm{He\,
    II}\right)+n\left(\mathrm{He\, III}\right)\nonumber \\ 
  =\frac{\rho\, Y}{\left(X+4Y\right)m_{p}}, 
\label{eq:he_num_cons}
\end{eqnarray}
where the numbers of energy levels used here are $N_{\mathrm{HeI}}=19$
and $N_{\mathrm{HeII}}=10$. Unless specified otherwise, we use $X=0.9$
and $Y=0.1$ for the abundances. Finally, the charge conservation
gives 
\begin{equation}
  n\left(\mathrm{H\,
    II}\right)+\sum_{i=1}^{N_{\mathrm{HeII}}}n_{i}\left(\mathrm{He\,
    II}\right)+2n\left(\mathrm{He\,
    III}\right)=n_{e}\,\,.
  \label{eq:q_cons}
\end{equation} 
Equations (\ref{eq:stateq}), (\ref{eq:hyd_num_cons}), (\ref{eq:he_num_cons})
and (\ref{eq:q_cons}) form a set of algebraic equations ($N_{\mathrm{HI}}+N_{\mathrm{HeI}}+N_{\mathrm{HeII}}+3$
equations) which contain $N_{\mathrm{HI}}+N_{\mathrm{HeI}}+N_{\mathrm{HeII}}+3$
unknowns. The unknowns here are the level populations of \ion{H}{I},
\ion{He}{I}, \ion{He}{II} respectively, and the last 3 unknowns
are the continuum levels and the electron density, i.e.~$n\left(\mathrm{H\, II}\right),$
$n\left(\mathrm{He\, III}\right)$ and $n_{e}$. The set of non-linear
algebraic equations will be solved by using the Newton-Raphson iteration
method. As a starting point on the iteration, we use either (1)~a
LTE population or (2) a set of non-LTE solutions from a nearby point
which had converged prior to the current computation. The level populations
normally converge within 30 iterations. 

As mentioned earlier, in our models presented here, the gas temperature
$T$ is given as an input because the exact physical processes that
determine the plasma temperature of the accretion flow and outflow
around CTTS (within a few stellar radii) are not well understood. We 
therefore hold the temperatures fixed and do not solve the equation
of radiative equilibrium.

\subsection{Observed profile calculations}

\label{sub:obs-prof}

The method for the profile calculation is essentially the same as
in \citet{kurosawa:2006}. Here, we briefly summarise the steps. Once
the level populations of H and He are obtained (as in Section~\ref{sub:stateq}),
the emissivity and opacities for the continuum and line ($\eta_{c}$,
$\chi_{c}$, $\eta_{l}$ and $\chi_{l}$) can be readily obtained
(e.g.~equation~\ref{eq:chi-sob}). By following \citet{mihalas:1978},
the total source function ($S_{\nu}$) can be written in terms of
the continuum and line source function which are defined as $S_{c}\equiv\eta_{c}/\chi_{c}$
and $S_{l}=\eta_{l}/\chi_{l}$, respectively, i.e. 
\begin{equation}
  S_{\nu}=\frac{\phi_{\nu}S_{l}+h\,
    S_{c}}{\phi_{\nu}+h} , 
  \label{eq:source-total}
\end{equation} 
where $h=\chi_{c}/\chi_{l}$ and $\phi_{\nu}$ is a normalized line
profile. When a line broadening is negligible $\phi_{\nu}$ is a normal
Doppler profile. However, as noted and demonstrated by \citet{muzerolle:2001}
and \citet{kurosawa:2006}, with a moderate mass-accretion rate ($~10^{-7}\,\mathrm{M_{\sun}\, yr^{-1}}$)
of a CTTS, Stark broadening becomes important in the optically thick
lines, e.g.~in H$\alpha$ and possibly in \ion{He}{i}~$\lambda10830$.
For those lines, we adopt a Voigt profile to include the broadening
effect. The normalized Voigt profile is written as $\phi_{\nu}=\pi^{-1/2}\, H\left(a,y\right)$
where 
\begin{equation}
  H\left(a,y\right)\equiv\frac{a}{\pi}\int_{-\infty}^{\infty}\frac{e^{-y'^{2}}}{\left(y-y'\right)^{2}+a^{2}}\,
  dy'\,\,.
  \label{eq:voigt_profile_def}
\end{equation} 
In the expression above, $a=\Gamma/4\pi\Delta\nu_{\mathrm{D}}$, $y=\left(\nu-\nu_{0}\right)/\Delta\nu_{\mathrm{D}}$,
and $y'=\left(\nu'-\nu_{0}\right)/\Delta\nu_{\mathrm{D}}$ (c.f. \citealt{mihalas:1978}).
$\nu_{0}$ is the line centre frequency, and $\Delta\nu_{\mathrm{D}}$
is the Doppler line width of an atom (due to its thermal motion) which
is given by $\Delta\nu_{\mathrm{D}}=\left(2kT/m_{\mathrm{i}}\right)^{1/2}\times\left(\nu_{0}/c\right)$
where $m_{\mathrm{i}}$ is either the mass of a hydrogen atom ($m_{\mathrm{H}}$)
or that of a helium atom ($m_{\mathrm{He}}$). The damping constant
$\Gamma$, which depends on the physical condition of the gas, is
parametrised by \citet{luttermoser:1992} as follows: 
\begin{eqnarray}
  \Gamma & = &
  C_{\mathrm{rad}}+C_{\mathrm{vdW}}
  \left(\frac{n_{\mathrm{HI}}}{10^{16}\,\mathrm{cm^{-3}}}\right)\left(\frac{T}{5,000\,\mathrm{K}}\right)^{0.3}\nonumber  \\
  &  & + \,C_{\mathrm{Stark}}\left(\frac{n_{e}}{10^{12}\,\mathrm{cm^{-3}}}\right)\,,
  \label{eq:dampimg_constant_def}
\end{eqnarray}
 where $n_{\mathrm{H\, I}}$ and $n_{e}$ are the number density of
neutral hydrogen atoms and that of free electrons. Also, $C_{\mathrm{rad}}$,
$C_{\mathrm{vdW}}$ and $C_{\mathrm{Stark}}$ are natural broadening,
van der Waals broadening, and linear Stark broadening constants respectively.
We adopt this parametrisation along with the values of broadening
constants from \citet{luttermoser:1992}. For example, $C_{\mathrm{rad}}=8.2\times10^{-3}$~\AA{},
$C_{\mathrm{vdW}}=5.5\times10^{-3}$~\AA{}~ and $C_{\mathrm{Stark}}=1.47\times10^{-2}$~\AA{}
for H$\alpha$. 

To find an observed line profile, we compute the flux as a function
of frequency. The integration of the flux is performed in the cylindrical
coordinate system $\left(p,\, q,\, t\right)$ which is obtained by
rotating the original stellar coordinate system $\left(\rho,\,\phi,\, z\right)$
so that the $t$-axis coincides with the line of sight of an observer.
The observed flux ($F_{\nu}$) is then given by: 
\begin{equation}
  F_{\nu}=\frac{1}{4\pi
    d^{2}}\int_{0}^{p_{\mathrm{max}}}\int_{0}^{2\pi}p\,\sin q\,
  I_{\nu} \, dq\,dp
  \label{eq:flux_integral}
\end{equation}
 where $d$, $p_{\mathrm{max}}$, and $I_{\nu}$ are the distance
to an observer, the maximum extent to the model space in the projected
(rotated) plane, and the specific intensity ($I_{\nu}$) in the direction
on observer at the outer boundary. For a given ray along $t$ or
alternatively using the optical depth ($\tau_{\nu}$ defined later in
equation~\ref{eq:optical-depth}) as an integration variable, the
specific intensity is given by: 
\begin{equation}
  I_{\nu}=I_{0}\,e^{-\tau_{\infty}}
   +\int_{0}^{\tau_{\infty}}S_{\nu}\left(\tau_{\nu}'\right)\,
  e^{-\tau_{\nu}'}\,d\tau_{\nu}'
  \label{eq:formal_sol_integral}
\end{equation}
 where $I_{0}$ and $S_{\nu}$ are the intensity at the boundary and the
 source function (equation~\ref{eq:source-total}) 
of the stellar atmosphere/wind at a frequency $\nu$. The upper
integration limit $\tau_{\infty}$ is the total optical depth between
an observer and the initial integration point. For a ray which
intersects with the stellar core, $I_{0}$ is computed from the stellar
atmosphere model of \citet{kurucz:1979} (cf.~Section~\ref{sub:cont-photosphere}),
and $I_{0}=0$ otherwise. If the ray intersects with the hot ring
on the stellar surface created by the accretion stream, we set $I_{0}=B_{\nu}\left(T_{\mathrm{ring}}\right)$
where $B_{\nu}$ is the Planck function and $T_{\mathrm{ring}}$ is
the temperature of the hot ring. The initial position of each ray
is assigned to be at the centre of the surface element ($dA=p\,\sin q\,dq\,dp$).
We numerically integrate equation~(\ref{eq:flux_integral}). A linearly
spaced radial grid is used for the area where the ray intersects with
magnetosphere, and a logarithmically spaced grid is used for the wind
and the accretion disc regions. The numbers of grid points used for
a typical integration are $n_{p}=180$, $n_{q}=100$, and $n_{\nu}=101$.
The optical depth $\tau_{\nu}$ in equation~\ref{eq:formal_sol_integral}
is defined as: 
\begin{equation}
  \tau_{\nu}\left(t\right)\equiv\int_{t}^{t_{\infty}}\chi_{\nu}\left(t'\right)\,dt'
  \label{eq:optical-depth}
\end{equation}
 where $\chi_{\nu}$ is the opacity of media, and $t_{\infty}$
is the total distance between the initial integration point
and the observer (or to the outer boundary closer to the observer).
Equation~(\ref{eq:optical-depth}) is also numerically integrated by using
the intersection points of a line along $t$ and the AMR grid cells,
in which opacities values ($\chi_{\nu}$) are stored. For a high optical
depth, additional points are inserted between the original points
along the line, and emissivity and opacity are interpolated to those
points to ensure $d\tau_{\nu}<0.05$ for the all line segments.

\section{Code tests}

\label{sec:results}

\begin{figure*}
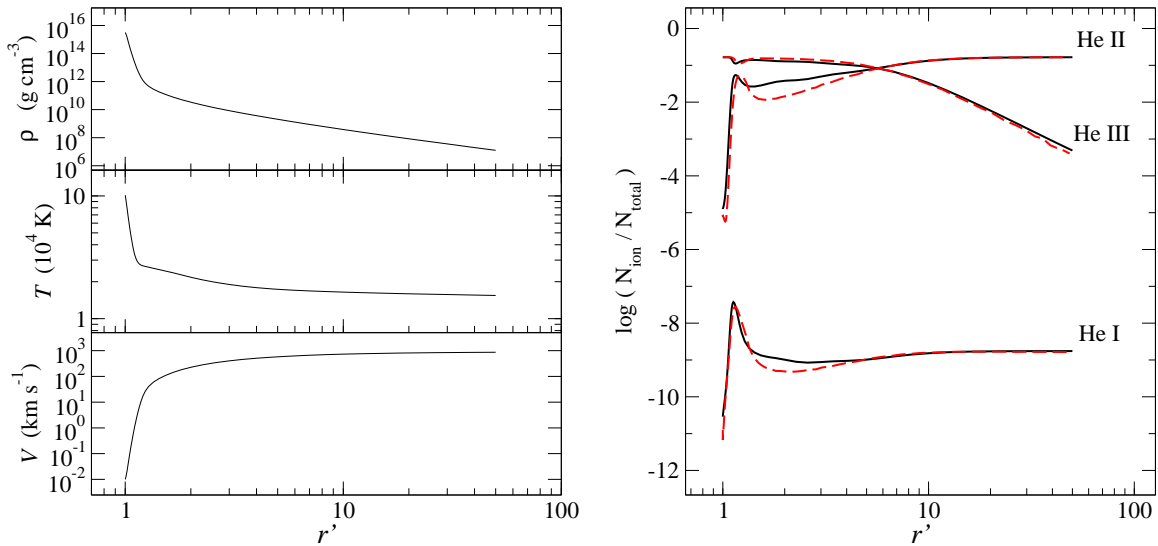



\begin{center}
    \begin{tabular}{c c}
      \includegraphics[clip,height=0.3\textheight]{fig02a.eps} & 
      \includegraphics[clip,height=0.3\textheight]{fig02b.eps} \\
    \end{tabular}
    \par
\end{center}

\caption{\emph{Left panel}:
  The density (top), temperature (middle) and radial
  velocity (bottom) of the hot stellar wind model of AV 83 computed by
  \textsc{cmfgen}. The values are shown as a function of radius $r'$, 
  which is in the unit of stellar radius (19.6\,$\Rsun$). The same
  density, temperature and velocity distributions are used in
  \textsc{torus} for the calculation of the statistical equilibrium. 
\emph{Right panel}: Comparison of the number fractions of helium ions as a
  function of radius, computed by {\sc cmfgen} (solid) and {\sc torus}
  (dashed). The fractions here are defined as
  $N_{\mathrm{ion}}/N_{\mathrm{total}}$ where $N_{\mathrm{ion}}$  is
  the number density of helium ions (\ion{He}{i}, \ion{He}{ii} or
  \ion{He}{iii}), and $N_{\mathrm{total}}$ is the total number density
  of ions, including both hydrogen and helium.  The fraction of
  \ion{He}{i} is much smaller compared to those of 
  \ion{He}{ii} and \ion{He}{iii}. Overall agreements of the two codes
  are very good.}

\label{fig:ion-frac}

\end{figure*}


\subsection{1D test : comparison with {\sc cmfgen} model of \citet{hillier:1998}}

\label{sub:test-1D}

We first test the validity of our statistical equilibrium calculation
and the observed profile calculations in a simple geometry. For this
purpose, we consider a case of spherical stellar wind from an Of supergiant
star since it often exhibits hydrogen and helium in emission. A detail
atmosphere model of AV~83 (an O7~Iaf+ supergiant) in the Small Magellanic
Cloud (SMC) has been presented by \citet{Hillier:2003}. Their models
are computed by the non-LTE line-blanketed atmosphere code {\sc cmfgen}
(\citealt{hillier:1998}) in which the level populations are solved
in the co-moving frame and in a spherical geometry. This code has
been well-tested and used in numerous investigations of stellar winds
of massive stars (e.g.~\citealt{hillier:1999}; \citealt{Crowther:1999};
\citealt{DeMarco:2000}; \citealt{Dessart:2000}; \citealt{herald:2001};
\citealt*{Martins:2005}; \citealt{Hillier:2006}). AV~83 is one of
two O stars in the SMC which shows a normal Of spectrum (c.f.~\citealt{Walborn:1995};
\citealt{Walborn:2000}). It has a relatively dense stellar wind,
and consequently exhibits many wind features in their spectrum. The
model presented by \citet{Hillier:2003} suggests that AV~83 has
the effective temperature of $\sim$32,800~K and the mass-loss rate
of $7.3\times10^{-7}\,\MsunPerYear$. 

In the following, we compare the results of the statistical equilibrium
calculations and the line profiles computed by {\sc torus} and {\sc cmfgen}.
Both models assume the hydrogen and helium abundances $X=0.834$ and
$Y=0.167$ (by numbers), respectively. To match the same physical
conditions in the stellar wind structure, {\sc torus} adopts the
temperature, density and velocity of the wind used in {\sc cmfgen}, 
which are shown in Fig.~\ref{fig:ion-frac}.
We also adopt the continuum flux from {\sc cmfgen} in computing the 
radiation fields in equations~\ref{eq:i-continuum} and \ref{eq:j-simple}. 
Further, we consider the case without the wind clumping, i.e. a smooth
stellar wind case. The {\sc cmfgen} model data of AV~83 used here
for comparison are available on the web site with the address: 
http://kookaburra.phyast.pitt.edu/hillier/web/CMFGEN.htm.  

Since our code {\sc torus} has newly adopted helium ions in the statistical
equilibrium calculation, here we concentrate on the comparisons of
the helium calculations. The hydrogen level population calculations
had been already tested elsewhere (\citealt{Harries:1995}; \citealt{harries:2000};
\citealt{symington:2005}). Fig.~\ref{fig:ion-frac} shows the number
fractions of helium ions in three stages (\ion{He}{i}, \ion{He}{ii}
and \ion{He}{iii}) as a function of radial distance from the centre
of the star ($r'$). The dominating ions of helium are \ion{He}{ii}
and \ion{He}{iii}, and the number fraction of \ion{He}{i} typically
$\sim10^{4}$ to $\sim10^{5}$ times smaller than those of other ions\@.
The figure shows the agreement of the ion fractions computed by both
codes is good at all radii, except at $r'\sim2$ (measured in stellar
radius) where \ion{He}{i} and \ion{He}{ii} fractions of {\sc torus}
are smaller than those of {\sc cmfgen} by a factor of a few. 

As one can see in Fig.~\ref{fig:ion-frac}, in this particular model
of AV~83, \ion{He}{ii} is the most abundant ion up to a several
radii. Correspondingly the optical thickness of the continuum photon
beyond the \ion{He}{ii} Lyman continuum is expected to be very high.
This contradicts with our assumption made in Section~\ref{sub:stateq}
which stated that the gas is optically thin to the photospheric continuum
radiation. Therefore, when the \ion{He}{ii} Lyman continuum is optically
thick, we introduce the following special treatment in the statistical
equilibrium. We assume that the radiation field beyond the \ion{He}{ii}
Lyman edge will be reduced to the local source function, and there
is radiative detail balance in the \ion{He}{ii} Lyman continuum.
In practice, we simply set the photorecombination and photoionization
terms in equations~\ref{eq:rate-recombine} and \ref{eq:rate-ionize}
for the ground state of \ion{He}{i} to negligibly small values. The
atmosphere is still assumed to be optically thin to continuum in all
other wavelengths. The same special treatment of the \ion{He}{ii}
Lyman continuum was used by KC78 who modelled the winds from Of stars.
Without this treatment, our ionization fraction of helium ions will
be very different from those of the {\sc cmfgen}  model at small
radii (e.g.~ $r'<2$) .

\begin{figure}

\begin{center}
\includegraphics[clip,width=0.48\textwidth]{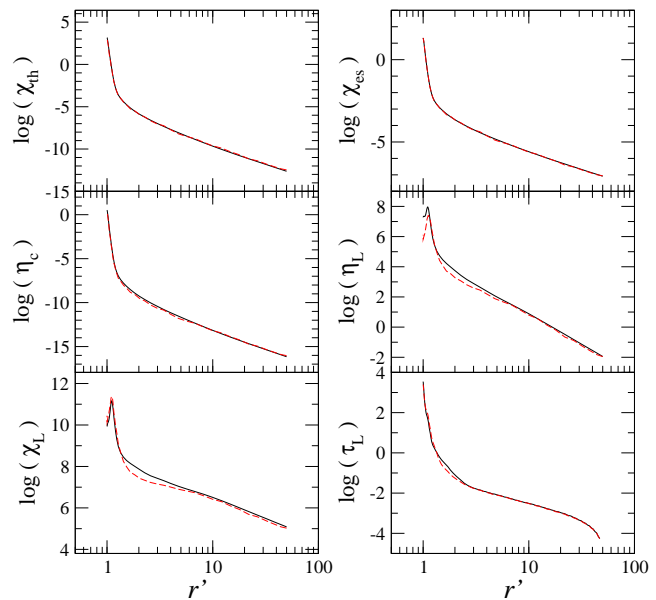}
\par\end{center}

\caption{
  Comparison of opacities, emissivities and optical depth as a function of
  radius, computed by {\sc cmfgen} (solid) and {\sc torus} (dashed).
  The line opacity ($\chi_{\mathrm{L}}$) and emissivity
  ($\eta_{\mathrm{L}}$)  are those of \ion{He}{i}~$\lambda$10830.  The
  continuum emissivity ($\eta_{\mathrm{c}}$) and the thermal opacity
  ($\chi_{\mathrm{th}}$) are evaluated at the line frequency of
  \ion{He}{i}~$\lambda$10830.  The electron scattering opacity is
  shown as $\chi_{\mathrm{es}}$.  The opacities and emissivities 
  shown are in arbitrary units. 
  The optical depth ($\tau_{L}$) is evaluated at the
  line centre of \ion{He}{i}~$\lambda$10830, and it is the sum of 
  the line and continuum optical depths.
  The radius
  ($r'$) is in the unit of stellar radius which is  19.6\,$\Rsun$.
  The agreements between the two codes are excellent for the continuum
  emissivity and opacities at all radii.  The agreements are also good
  in the line opacity and emissivity except for those at the inner
  most radii ($r'<1.1$) where the optical depth is very high and the
  outflow speed is subsonic. } 

\label{fig:opacity}

\end{figure}

Fig.~\ref{fig:opacity} compares the line emissivity and opacity
of \ion{He}{i}~$\lambda$10830 computed by {\sc torus} and {\sc cmfgen}
as a function of radius. As mentioned earlier, this particular line
has been identified as one of the important wind diagnostic lines
in CTTSs (e.g.~\citealt{Edwards:2006}; \citealt{Kwan:2007}). The
figure also shows the electron scattering opacity, and the continuum
opacity and emissivity at the line centre of \ion{He}{i}~$\lambda$10830.
Overall agreement between the two codes is excellent for the continuum
emissivity and opacities at all radii.  The agreements are also good
in the line opacity and emissivity except for the line emissivity
($\eta_{\mathrm{L}}$) at the inner most radii ($r'<1.1$) where the
outflow speed is subsonic; hence, the Sobolev approximation is not
suitable there. 
The figure also shows the corresponding optical depth
  ($\tau_{L}$) evaluated at the line centre of
  \ion{He}{i}~$\lambda$10830.


\begin{figure}

  \begin{center}
    \includegraphics[clip,width=0.48\textwidth]{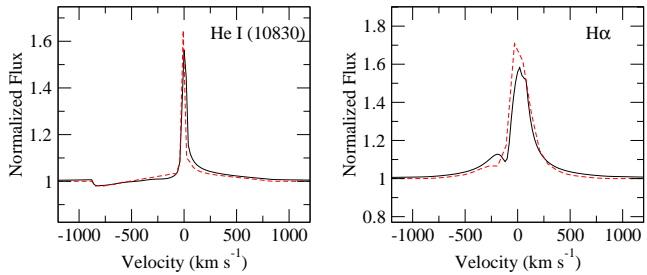}
  \end{center}

  \caption{Comparison of line profiles (\ion{He}{i}~$\lambda10830$: right
    panel and  H$\alpha$: left panel) computed by {\sc cmfgen} (solid)
    and {\sc torus} (dashed).  The line strengths of the profiles
    computed by {\sc torus} are slightly stronger than those of {\sc
      cmfgen}; however, overall agreement of the models is excellent.
    In both lines, the broadening mechanisms such as the Stark
    broadening and the rotational broadening are not included. } 
  \label{fig:test1d-profile}

\end{figure}

Using the opacities and emissivities above, we now compute the line
profiles as seen by an observer. Fig.~\ref{fig:test1d-profile} shows
the normalized profile of \ion{He}{i}~$\lambda$10830 and H$\alpha$
computed by {\sc torus} and {\sc cmfgen}. For a simplicity, the
broadening effects such as the Stark broadening and rotational broadening
are not included in these calculations. \ion{He}{i}~$\lambda$10830
shows a narrow emission near the line centre and a shallow but very
wide wind absorption in the blue side. On the other hand, H$\alpha$
profiles show a wider emission and a relatively narrow wind absorption
in the blue. The extent of the red wings in the \ion{He}{i}~$\lambda$10830
and H$\alpha$ lines is similar, and they both reach the velocity $\sim800\,\kmps$
which is very similar to the terminal velocity of the stellar wind
($900\,\kmps$) used in the models. The figure shows the emission
components of both H$\alpha$ and \ion{He}{i}~$\lambda$10830 profiles
computed by {\sc torus} are slightly stronger than those by {\sc
  cmfgen}; however, overall agreement of the line profiles with those
of {\sc cmfgen} is very good.


\subsection{2D test: comparison with models from \citet{muzerolle:2001}}

\label{sub:test-2D}

Next, we examine a slightly more complex geometry, namely the axisymmetric
magnetospheric accretion flows on to the CTTS as described by \citet{hartmann:1994}.
In this model, the gas accretion on to the stellar surface from the
innermost part of the accretion disc occurs through a bipolar stellar
magnetic field. The magnetic field and the gas stream lines are assumed
to have the following simple form 
\begin{equation}
  r=r_{\mathrm{m}}\sin^{2}\theta
  \label{eq:dipole}
\end{equation} 
(cf.~\citealt*{ghosh:1977}) where $r$ and $\theta$ are the radial
and the polar components of a position vector in the accretion stream.
The symbol $r_{\mathrm{m}}$ is the radial distance to the field line
on the equatorial plane ($\theta=\pi/2$), and its value is restricted
to be between $r_{\mathrm{in}}$ and $r_{\mathrm{out}}$ (cf.~Fig.~1 in
\citealt{hartmann:1994}) which control the size and width of the magnetosphere.
Using the field geometry above and conservation of energy, the velocity
and the density of the accreting gas along the stream line are found
as in \citet{hartmann:1994}. We also adopt the temperature structure
along the stream lines used by \citet{hartmann:1994}. In fact,
the flow geometry and the density structure described here are used
to construct the AMR grid cells shown in Fig.~\ref{fig:grid-example}
(Section~\ref{sub:grid-construction}). The same magnetospheric accretion
structure is used in the radiative transfer models by \citet{muzerolle:2001},
\citet{symington:2005} and \citet{kurosawa:2006}. 

\citet{muzerolle:2001} presented hydrogen line profiles of CTTS with
many different combinations of magnetosphere geometries, flow temperatures
and mass-accretion rates. As our test, we consider one representative
model of \citet{muzerolle:2001}, and compute line profiles to check
if we can reproduce their results. The following model configuration
is used in this test. The mass ($M_{*}$) and radius ($R_{*}$) of
the central star are assumed to be 0.5~$\Msun$ and 2~$\Rsun$,
respectively. The effective temperature of the star is set to 4,000~K,
and the maximum temperature in the magnetosphere ($T_{\max}$ in \citealt{hartmann:1994})
is 7,000~K. The geometry of the magnetosphere is determined by setting
$r_{m}$ to $r_{\mathrm{in}}=2.2\, R_{*}$ and $r_{\mathrm{out}}=3.0\,
R_{*}$ in equation~\ref{eq:dipole}. This corresponds to the {}``small/wide''
model in \citet{muzerolle:2001}. Finally, the mass-accretion rate
is set to $10^{-7}\,\MsunPerYear$. 

\begin{figure}

\begin{center}
\includegraphics[clip,width=0.48\textwidth]{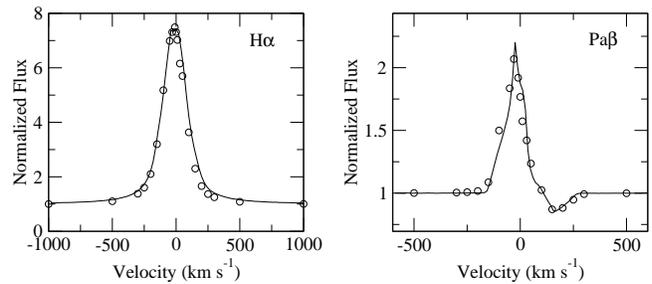}
\par\end{center}

\caption{Comparison of H$\alpha$ (left) and Pa$\beta$ (right) profiles
  for an axisymmetric accretion flow as described by
  \citet{hartmann:1994}. The profiles computed by   {\sc torus} (solid
  line)  are compared with those from \citet{muzerolle:2001}
  (circles). The mass and radius of the central star are
  $M_{*}=0.5\,\Msun$ and $R_{*}=2\,\Rsun$, respectively. The effective
  temperature of the star is 4,000\,K.  The accretion stream has and
  the maximum temperature  $T_{\mathrm{max}}$=7,000K, and the
  mass-accretion rate $\dot{M}_{\mathrm{a}}=10^{-7}\,\MsunPerYear$.
  The accretion stream starts in the disc plane between the radial
  distances $r_{\mathrm{in}}=2.2\,R_{*}$  and
  $r_{\mathrm{in}}=3.0\,R_{*}$ . The star is viewed at the inclination
  angle of $60^{\circ}$.} 

\label{fig:compare-mch}

\end{figure}

Fig.~\ref{fig:compare-mch} shows H$\alpha$ and Pa$\beta$ profiles
computed with this model setup. The figure also shows the profile
computed by \citet{muzerolle:2001} for comparison. In computing the
H$\alpha$ profile, we have taken into account the line broadening
effect as described in section~\ref{sub:obs-prof}. Although the
line wings of our H$\alpha$ profiles are slightly wider than those
of \citet{muzerolle:2001}, overall agreement between our profiles
and their model profiles is very good. A similar test can be also
found in \citet{symington:2005}. 

We are not able to compare our model \ion{He}{i} line profiles with
the models of \citet{muzerolle:2001} since their model does not include
helium atoms. In fact, \ion{He}{i} lines would not be formed with
this model configuration because of the low temperature in the accretion
stream and/or the lack of high energy (photoionization) radiation source
which could excite or ionize helium atoms. The importance of the photoionization
by high energy photons will be discussed further in Sections \ref{sub:cont-xray},
\ref{sub:ma-dw-example} and \ref{sub:ma-sw-example}.

\section{Applications to classical T Tauri Stars}

\label{sec:apps-to-ctts}

We now apply the radiative transfer model, described and tested earlier,
to more comprehensive flow configurations of CTTSs which include both
a magnetospheric accretion and a wind. In the following, we briefly
describe our model configurations including the sources of the continuum
radiation, the disc wind model and the stellar wind model. The example
profiles are computed with the combinations of (1)~magnetospheric
accretion and disc wind, and (2)~magnetospheric accretion and the
stellar wind. Here, we consider only the axisymmetric cases. In both
cases (1) and (2), the magnetospheric accretion model of \citet{hartmann:1994}
is adopted, which is briefly described in Section~\ref{sub:test-2D}
and its accretion geometry is shown in Fig.~\ref{fig:grid-example}.

\subsection{Continuum sources}

\label{sub:continuum_sources}

\subsubsection{Photosphere}

\label{sub:cont-photosphere}

We adopt stellar parameters of a typical classical T~Tauri star for
the central continuum source, i.e.~ its stellar radius $R_{*}=2.0\,\Rsun$
and its mass $M_{*}=0.5\,\Msun$. Consequently, we adopt the effective
temperature of photosphere $T_{\mathrm{ph}}=4,000\,\Kelvin$ and the surface
gravity $\log g_{*}=3.5$ (cgs), and use the model atmosphere of \citet{kurucz:1979}
as the photospheric contribution to the continuum flux.

\subsubsection{Hot spots/rings}

\label{sub:cont-hotspot}

Additional continuum sources to be included are the hot spots/rings
formed by the infalling gas along the magnetic field on to the stellar
surface. As the gas approaches the surface, it decelerates in a strong
shock, and is heated to $\sim10^{6}\,\Kelvin$. The X-ray radiation
produced in the shock will be absorbed by the gas locally, and re-emitted
as optical and UV light (\citealt{calvet:1998}; \citealt{gullbring:2000})
-- forming the high temperature regions on the stellar surface with
which the magnetic field intersects. In this study, we adopt a single
temperature model for hot spots/rings by assuming the free-falling
kinetic energy is thermalised in the radiating layer, and is re-emitted
as blackbody radiation, as described by \citet{hartmann:1994}. The
same model is used in the radiative transfer models by \citet{muzerolle:2001},
\citet{symington:2005} and \citet{kurosawa:2006}. The hot spot/ring
temperature ($T_{\mathrm{hs}}$) in this model is summarised as: 
\begin{equation}
  T_{\mathrm{hs}}^{4}=\frac{L_{\mathrm{hs}}}{4\pi
    R_{*}^{2}\sigma\left(\cos\theta_{\mathrm{in}}-\cos\theta_{\mathrm{out}}\right)}\,,
  \label{eq:hotspot-temperature}
\end{equation} 
 where $L_{\mathrm{hs}}$ and $\sigma$ are the hot spot/ring luminosity
and the Stefan-Boltzmann constant. The angles $\theta_{\mathrm{in}}$
and $\theta_{\mathrm{out}}$ represent the positions (the polar angles)
where the innermost and outermost dipolar magnetic field line intersects
with the stellar surface (cf.~Figs~\ref{fig:grid-example}, \ref{fig:config-MA-DW},
\ref{fig:config-MA-SW}). Using equation~(\ref{eq:dipole}) along with the inner
and outer radii of the magnetosphere ($R_{\mathrm{mi}}$ and $R_{\mathrm{mo}}$,
respectively), we find $\theta_{\mathrm{in}}=\sin^{-1}\left[\left(R_{*}/R_{\mathrm{mi}}\right)\right]^{1/2}$
and $\theta_{\mathrm{out}}=\sin^{-1}\left[\left(R_{*}/R_{\mathrm{mo}}\right)\right]^{1/2}$.
The hot spot/ring luminosity in this model can be written as 
\begin{equation}
  L_{\mathrm{hs}}=\frac{GM_{*}\dot{M}_{\mathrm{a}}}{R_{*}}\left[1-\frac{2R_{*}}{\left(R_{\mathrm{mi}}+R_{\mathrm{mo}}\right)}\right]\,,
  \label{eq:hotspot-luminosity}
\end{equation} 
 where $\dot{M}_{\mathrm{a}}$ is the mass-accretion rate. Readers
are referred to \citet{hartmann:1994} for details. With the stellar
parameters used in Section~\ref{sub:cont-photosphere} along with
$\dot{M}_{\mathrm{a}}=10^{-7}\,\MsunPerYear$, $R_{\mathrm{mi}}=2.2\, R_{*}$
and $R_{\mathrm{mo}}=3.0\, R_{\mathrm{*}}$, the corresponding hot
spot/ring temperature and the accretion luminosity are $T_{\mathrm{hs}}=6,400\,\Kelvin$
and $L_{\mathrm{hs}}=1.9\times10^{33}\,\mathrm{erg\, s^{-1}}=0.49\,\mathrm{L_{\sun}}$. 

Alternatively, we can also construct the multi-temperature hot spot
model obtained in the 3D MHD simulations by \citet{romanova:2004} in which the local hot spot
luminosity is computed directly from the inflowing mass flux on the
surface of the star. In this model, the temperature of the hot spots
is determined by conversion of kinetic plus internal energy
of infalling gas to a blackbody radiation. They obtained
the position-dependent mass flux crossing the inner boundary; hence,
achieving a position-dependent temperature of the hot spot, which
can be written as
\begin{equation}
  T_{\mathrm{hs}}=\left\{
  \frac{\rho\,\left|v_{r}\right|}{\sigma}\left(\frac{1}{2}v^{2}+w\right)^{2}\right\}^{1/4}\,,
  \label{eq:hot_spot_romanova}
\end{equation}
 where $\rho$, $v_{r}$ and $v$ are the density, the radial component
of velocity and the speed of gas/plasma, respectively. The specific
enthalpy of the gas is $w=\gamma\left(p/\rho\right)\left(\gamma-1\right)$
where $\gamma$ is the adiabatic index and $p$ is the gas pressure.
This model is useful when the accretion flow from a MHD simulation
(e.g.~\citealt{romanova:2003}; \citealt{Romanova:2009}) is used,
as in \citet{kurosawa:2008}.

We compare this temperature $T_{\mathrm{hs}}$ with the effective
temperature of photosphere $T_{\mathrm{ph}}$ to determine the shape
and the size of hot spots. When $T_{\mathrm{hs}}>T_{\mathrm{ph}}$,
the location on the stellar surface is flagged as hot. For the hot
surface, the total continuum flux is the sum of the blackbody radiation
with $T_{\mathrm{hs}}$ and the flux from the model photosphere mentioned
above. The contribution from the inflow gas is ignored when $T_{\mathrm{hs}}<T_{\mathrm{ph}}$.

\subsubsection{X-ray emission}

\label{sub:cont-xray}

CTTSs show relatively strong X-ray emission with their luminosities
$L_{\mathrm{X}}\left(0.3-10\,\mathrm{keV}\right)$ ranging from $\sim10^{28}\,\mathrm{erg\, s^{-1}}$
to $\sim10^{31}\,\mathrm{erg\, s^{-1}}$ (e.g.~\citealt{Telleschi:2007};
\citealt{Gudel:2007}; \citealt{Gudel:2010}). The soft X-ray emission
component (with the electron temperature $T_{\mathrm{e}}\sim$a few
$10^{6}$~K) are possibly associated with the shock regions either
near the base of the accretion columns (e.g.~\citealt{calvet:1998};
\citealt{Lamzin:1998}; \citealt{Kastner:2002}; \citealt{Orlando:2010})
or the base of the stellar wind/jet (e.g.~\citealt{Gudel:2008};
\citealt{Schneider:2008}). On the other hand, the higher temperature
($T_{\mathrm{e}}\sim10\times10^{6}$~K) emission may be associated
with the flares and active coronal regions (e.g.~\citealt{Feigelson:1999};
\citealt*{Gagne:2004}; \citealt{Preibisch:2005}; \citealt{Jardine:2006};
\citealt{Stassun:2006}; \citealt*{Argiroffi:2007}). 

The coronal heating and the production of X-ray are closely related
to the formation and the acceleration of the stellar wind from CTTSs
(e.g.~\citealt{Cranmer:2009}). Exploring possible formation mechanisms
of the stellar wind is important because its mass-loss rate is directly
related to the amount of the angular momentum that can be removed
from the accreting CTTSs (e.g.~\citealt{matt:2005}; \citeyear{Matt:2007b};
\citeyear{Matt:2008a}). The X-ray irradiation on the accretion disc
is also important for its thermodynamics, structure and chemistry,
and it is the main source of ionization in the outer layers of the
disc (e.g.~\citealt*{glassgold:2004}; \citealt{Alexander:2008};
\citealt{Bergin:2009}; see also \citealt{Shang:2002} for the X-ray
heating of the X-wind). It can also assist a disc to lose its mass at
large radii via photoevapolation (e.g.~\citealt*{Ercolano:2009};
\citealt*{Owen:2010}). The importance of the X-ray radiation in determining
the physical conditions of the circumstellar material of CTTSs is
clear; however, so far, few spectroscopic models of optical and near-infrared
He and H lines that include the effect of photoionization due to X-ray
have been explored. For this reason, we will include the effect of the
X-ray radiation in our calculations.  Note that a recent local excitation model of
\citet{Kwan:2011} demonstrated that the photoionization of He due to
`UV radiation' might be also important for ionizing He. However, we
concentrate on X-ray radiation here, and leave the investigation
of the relative importance of X-ray and UV radiation as our future
study. 

Another important motivation for adding the X-ray source in our model
comes from our initial investigation of the simultaneous modelling of
H and He emission lines in the CTTSs. Although not shown here, we have
found that the temperature of the flows (both accretion and wind) must
be relatively low e.g.~$T\lesssim$10,000~K\footnote{See also Fig.~16
in \citet{muzerolle:2001} for the possible temperature ranges of the
axisymmetric magnetospheric accretion flow, based on their hydrogen
emission line profile models.}  for a system with the 
mass-accretion rate, and it could be slightly higher for a lower
mass-accretion rate. The line strengths of hydrogen lines (e.g.~H$\alpha$) will be
unrealistically strong if the temperature is much above this
temperature. However, at this relatively low temperature, the
collisional rates of \ion{He}{i} are low, and the normal continuum
sources mentioned earlier (Sections~\ref{sub:cont-photosphere} and
\ref{sub:cont-hotspot}) do not provide enough radiation to
ionize/excite \ion{He}{i} significantly. This results in no
significant emission or absorption in \ion{He}{i}~$\lambda10830$, and
is inconsistent with the observations
(e.g.~\citealt{Edwards:2006}). One way to overcome this problem is to
introduce an extra source of He ionizing radiation, e.g.~an X-ray
emitting source. The X-ray radiation will photoionize \ion{He}{i} and
keeps its excited states populated.

Here, we take a simplest approach in implementing the X-ray radiation
in our model. Firstly, we assume that the X-ray radiation arises uniformly
from the stellar surface as though it were formed in the
chromosphere.  This assumption allows us to include the X-ray emission
by simply adding it to the normal stellar continuum flux, discussed
in Section~\ref{sub:cont-photosphere}. The dependency of the line
formation on the different X-ray emission locations is beyond the
scope of this paper, but will be explored in the future. Secondly,
we assume that the energy distribution is flat in the energy range
between 0.1 to 10~keV, and the X-ray flux is normalized according
to the total X-ray luminosity $L_{\mathrm{X}}\left(0.1-10\,\mathrm{keV}\right)$,
which we set as an input parameter. Alternatively, we could adopt
a more realistic energy distribution similar to those used in the
X-ray photoevapolation disc model of \citet{Ercolano:2008}, who used
the synthetic X-ray spectra of a typical CTTS by assuming the coronal
X-ray emission is produced in a collision dominated optically thin
plasma. The attenuation of X-ray flux is not taken into account in our
model; however, this can be improved in a future study. 
A further discussion of the X-ray flux attenuation is given in
Section~\ref{sub:x-ray-tau}. 

Using the X-ray emission as assumed above, we find a typical value of
the \ion{He}{i} photoionization rate ($\gamma_{\mathrm{He\,I}}$) from
the ground state, at a typical distance from the star
($r=4.0\,R_{*}$), is $\gamma_{\mathrm{He\,I}}\sim 3\times
10^{-5}\,\mathrm{s^{-1}}$ when the X-ray luminosity is assumed to be
$L_{\mathrm{X}}=2\times 10^{30}\,\mathrm{erg\,s^{-1}}$.
Interestingly, this is very similar to the values used in the local
excitation calculations of \citet{Kwan:2011},
i.e.~$\gamma_{\mathrm{He\,I}}=10^{-4}$ and $10^{-5}\,\mathrm{s^{-1}}$.

\subsection{Disc wind model}

\label{sub:discwind-model}

\subsubsection{Configuration}

\label{sub:discwind-config}

The disc wind model used in this analysis is essentially same as in
\citet{kurosawa:2006}, who adopted a simple kinematical wind model
of \citet*{knigge:1995} and follows the basic ideas of the magneto-centrifugal
wind paradigm (e.g.~\citealt{blandford:1982}). The model is designed
to broadly represent the MHD wind models (e.g.~\citealt{shu:1994};
\citealt{ustyugova:1995}; \citealt{ouyed:1997};
\citealt{Romanova:1997}; 
\citealt{Ustyugova:1999};
\citealt{koenigl:2000};
\citealt{krasnopolsky:2003};  \citealt{Romanova:2009};
\citealt*{Murphy:2010}). In the following, we briefly describe our
disc wind model which is an adaptation of the {}``split-monopole''
wind model by \citet{knigge:1995}. In this model, the outflow arises
from the surface of the rotating accretion disc, and has a biconical
geometry. The specific angular momentum is assumed to be conserved
along a stream line, and the poloidal velocity component is assumed
to be simply a radial from {}``sources'' vertically displaced from
the central star. Readers are referred to \citet{knigge:1995} and
\citet{long:2002} for details. See \citet{alencar:2005} and
\citet{Lima:2010} for an alternative disc wind model.

\begin{figure}

\begin{center}

\includegraphics[clip,width=0.4\textwidth]{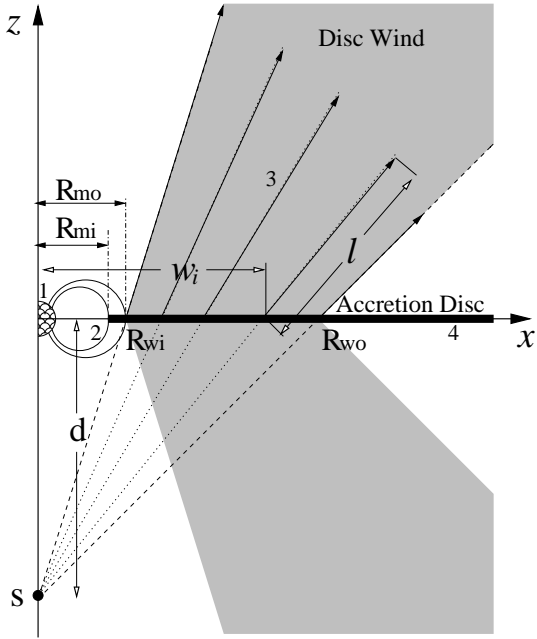}

\end{center}

\caption{Basic model configuration of the disc-wind-magnetosphere
  hybrid model. The system consists of four components: (1)~the
  continuum source located at the origin of the cartesian coordinates
  $\left(x,y,z\right)$ -- the $y$-axis is into the paper, (2)~the
  magnetospheric accretion flow, (3)~the disc wind, and (4)~the
  optically thick but geometrically thin accretion disc. The disc wind
  originates from the disc surface between $w_{i}=R_{\mathrm{wi}}$ and
  $R_{\mathrm{wo}}$ where $w_{i}$ is the distance from the $z$
  axis on the equatorial plane. The wind source points ($S$), from
  which the stream lines diverge, are placed at distance $d$ above and
  below the star. The degree of wind collimation is controlled by
  changing the value of $d$. The figure is not to scale.} 

\label{fig:config-MA-DW}

\end{figure}

There are four basic parameters: (1)~the mass-loss rate, (2)~the
degree of the wind collimation, (3)~the velocity gradient, and (4)~the
wind temperature. The basic configuration of the disc-wind model is
shown in Fig.~\ref{fig:config-MA-DW}. The disc wind originates from
the disc surface, but the {}``source'' point ($S$), from which
the stream lines diverge, are placed at distance $d$ above and below
the centre of the star. The angle of the mass-loss launching from
the disc is controlled by changing the value of $d$. The mass-loss
launching occurs between $R_{\mathrm{wi}}$ and $R_{\mathrm{wo}}$
where the former is set to be equal to the outer radius of the closed magnetosphere
($R_{\mathrm{mo}}$) and the latter is set to $0.5$~au.

The local mass-loss rate per unit area ($\dot{m}$) is assumed to
be proportional to the mid-plane temperature of the disc, and is a
function of the cylindrical radius $w=\left(x^{2}+y^{2}\right)^{1/2}$,
i.e. 
\begin{equation}
  \dot{m}\left(w\right)\propto\left[T\left(w\right)\right]^{4\alpha}\,\,.
  \label{eq:discwind-massloss-temp}
\end{equation} 
The mid-plane temperature of the disc is assumed to be expressed
as a power-law in $w$; thus, $T\propto w^{q}$. Using this in the
relation above, one finds 
\begin{equation}
  \dot{m}\left(w\right)\propto w^{p}\,,
  \label{eq:discwind-massloss-w}
\end{equation}
where $p=4\alpha\times q$. The index of the mid-plane temperature
power law is adopted from the dust radiative transfer model of \citet{whitney:2003a}
who found the innermost part of the accretion disc has $q=-1.15$. To
be consistent with the collimated disc-wind model of
\citet{krasnopolsky:2003} who used $p=-7/2$, the value of $\alpha$ is set to 0.76. The constant
of proportionality in equation~\ref{eq:discwind-massloss-w} is found
by integrating $\dot{m}$ from $R_{\mathrm{wi}}$ to $R_{\mathrm{wo}}$,
and normalising the value to the total mass-loss rate $\dot{M}_{\mathrm{dw}}$.

The azimuthal/rotational component of the wind velocity $v_{\phi}\left(w,z\right)$
is computed from the Keplerian rotational velocity at the emerging
point of the stream line i.e. $v_{\phi}\left(w_{i},0\right)=\left(GM_{*}/w_{i}\right)^{1/2}$
where $w_{i}$ is the distance from the rotational axis ($z$) to
the emerging point on the disc, and by assuming the conservation of
the specific angular momentum along a stream line: 
\begin{equation}
  v_{\phi}\left(w,z\right)\,=v_{\phi}\left(w_{i},0\right)\,\left(\frac{w_{i}}{w}\right)\,\,.
  \label{eq:discwind-toroidal-velocity}
\end{equation} 
Based on the canonical $\beta$ velocity law of hot stellar winds
(c.f.~\citealt*{castor:1975}), the poloidal component of the wind velocity
($v_{p}$) parametrised as: 
\begin{equation}
  v_{p}\left(w_{i},l\right)=c_{\mathrm{s}}\left(w_{i}\right)+\left[f\,
    v_{\mathrm{esc}}-c_{\mathrm{s}}\left(w_{i}\right)\right]\left(1-\frac{R_{s}}{l+R_{s}}\right)^{\beta}\,,
  \label{eq:discwind-poloidal-velocity}
\end{equation} 
where $c_{\mathrm{s}}$, $f$, and $l$ are the sound speed at the
wind launching point on the disc, the constant scale factor of the
asymptotic terminal velocity to the local escape velocity ($v_{\mathrm{esc}}$
from the wind emerging point on the disc), and the distance from the
disc surface along stream lines respectively. $R_{s}$ is the wind
scale length, and its value is set to 10~$R_{\mathrm{wi}}$ by following
\citet{long:2002}. We set $f=2$ in the models presented in this
paper.

Assuming mass-flux conservation and using the velocity field defined
above, the disc wind density as a function of $w_{i}$ and $l$ can
be written as 
\begin{equation}
  \rho\left(w_{i},l\right)=\frac{\dot{m}\left(w_{i}\right)}{v_{p}\left(w_{i},l\right)\,\left|\cos\delta\right|}\,\left\{
  \frac{d}{D\left(w_{i},l\right)\,\cos\delta}\right\} ^{2} \,,
  \label{eq:discwind-density}
\end{equation}
where $D$ and $\delta$ are the distance from the source point ($S$)
to a point along the stream line and the angle between the stream
line and the disc normal, respectively. The dependency of the density
and the poloidal velocity on the wind acceleration parameter $\beta$
is shown in Fig.~2 of \citet{kurosawa:2006}. Although the outer
radius ($R_{\mathrm{wo}}$) of the disc wind could extend to a large
radius, most of the mass-loss occurs near the inner radius ($R_{\mathrm{wi}}$)
of the disc wind because of the rather steep radial dependency of
the local mass-loss rate ($p=-7/2$ in equation~\ref{eq:discwind-massloss-w})
adopted here. This concentration of the mass-loss near the
disc-magnetosphere 
interaction region resembles that of the conical wind model of
\citet{Romanova:2009}. Further, by restricting the outer disc radius
($R_{\mathrm{wo}}$) to a smaller radius, our disc wind model can
mimic the conical wind model even better.  Lastly, the temperature of
the disc wind ($T_{\mathrm{dw}}$) is assumed isothermal as in
\citet{kurosawa:2006}.

\subsubsection{Example profiles}

\label{sub:ma-dw-example}

\begin{figure}

\includegraphics[clip,width=0.48\textwidth]{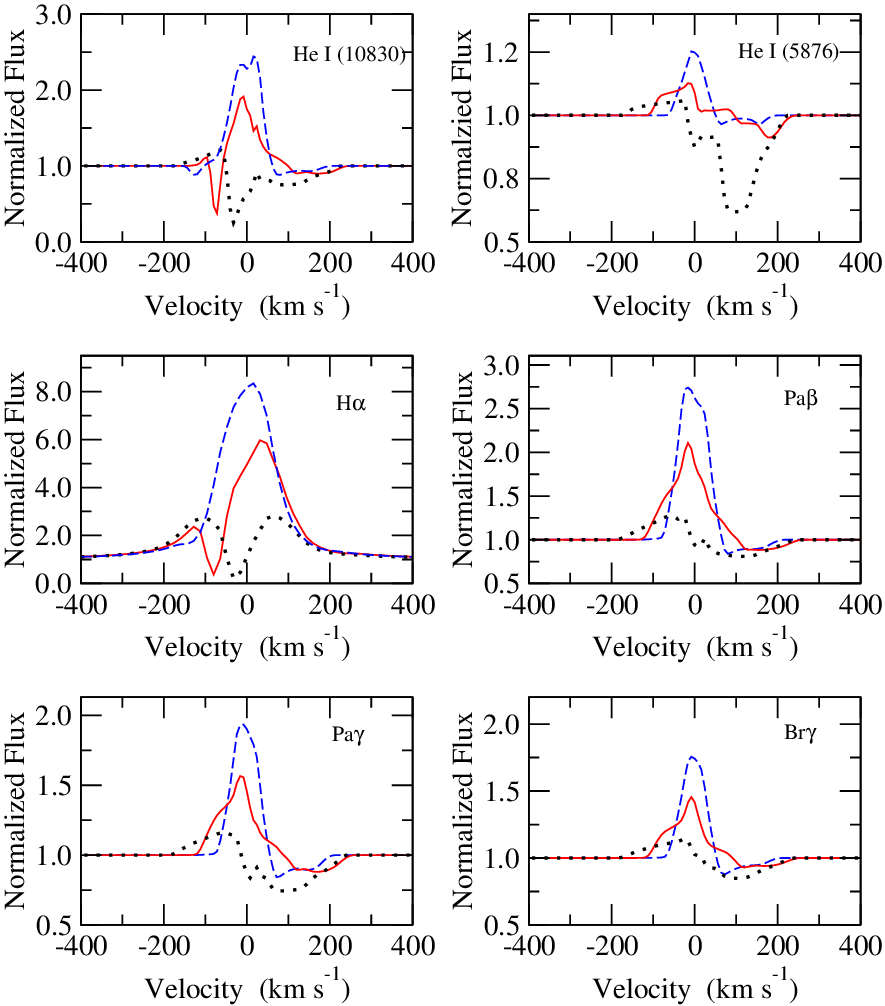}

\caption{An example model calculation with a combination of
  axisymmetric magnetosphere and disc wind. The normalized model
  profile are shown as a function of velocity. The model parameters
  adopted here are: $T_{\mathrm{max}}=$7,500~K,
  $T_{\mathrm{dw}}=$9,000~K,
  $L_{\mathrm{X}}=2\times10^{29}\,\mathrm{erg s^{-1}}$, and
  $\dot{M}_{\mathrm{a}}=10^{-7}\,\MsunPerYear$,
  $\dot{M}_{\mathrm{dw}}=10^{-8}\,\MsunPerYear$ (see text for other
  parameters).  The profiles are computed at inclination angles
  $i=20^{\circ}$ (dashed), $50^{\circ}$ (solid) and $80^{\circ}$
  (dotted).
    N.B. the lower limits of the vertical axis for \ion{He}{i}~(10830)
    and H$\alpha$ are set to 0, but those for \ion{He}{i}~(5876),
    Pa$\beta$, Pa$\gamma$ and Br$\gamma$ are set to 0.5.  
}   

\label{fig:example-MA-DW-all}

\end{figure}

Fig.~\ref{fig:example-MA-DW-all} shows the example helium and hydrogen
profiles computed for the standard axisymmetric magnetospheric accretion
funnels ($R_{\mathrm{mi}}=2.2\, R_{*}$, $R_{\mathrm{mo}}=3.0\, R_{*}$,
$\dot{M}_{\mathrm{a}}=10^{-7}\,\MsunPerYear$ and $T_{\mathrm{max}}=7,500\,\Kelvin$)
and the disc wind with the following parameters: $T_{\mathrm{dw}}=9,000\,\Kelvin$,
$\dot{M}_{\mathrm{dw}}=10^{-8}\,\MsunPerYear$, $\beta=0.5$, $R_{\mathrm{wi}}=3.0\, R_{*}$,
$R_{\mathrm{wo}}=54\, R_{*}$, $d=14\, R_{*}$ (cf.~Fig.~\ref{fig:config-MA-DW}).
The parameters for the central star are as described in Section~\ref{sub:cont-photosphere}.
The effective temperature of the hotspot, as described in Section~\ref{sub:cont-hotspot},
is approximately 6,400~K. The X-ray luminosity of the chromosphere
(cf.~Section~\ref{sub:cont-xray}) is assumed to be $L_{\mathrm{X}}=2\times10^{29}\,\mathrm{erg\, s^{-1}}$.
The profiles shown here are computed for the system with the inclination
angle of $i=45^{\circ}$. 
The stellar and disk occultations are taken into account in our
profile calculations.

We have adopted a relatively low disc wind temperature ($9,000\,\Kelvin$)
here to avoid unrealistically strong emission in the hydrogen lines,
produced at the base of the wind. For the same reason, the temperature
of the magnetosphere cannot be much higher than $T_{\mathrm{max}}=7,500\,\Kelvin$
(for the system with $\dot{M}_{\mathrm{a}}=10^{-7}\,\MsunPerYear$
or for the corresponding density in the accretion flow). As mentioned
earlier, at these relatively low kinetic temperatures of the gas,
the collisional rates of \ion{He}{i} are not strong enough to populate
the excited states; hence, an important wind diagnostic line such
as \ion{He}{i}~$\lambda$10830 will not be formed in emission nor
in absorption without an additional source of excitation/ionization
mechanism. For this reason, we have added the X-ray continuum (as
described in Section~ \ref{sub:cont-xray}) which can photoionize
\ion{He}{i}. We find that in general the line strength of \ion{He}{i}~$\lambda$10830
is sensitive to the amount of the X-ray luminosity ($L_{\mathrm{X}}$)
adopted in our model. 
The model profiles computed at
three different inclination angles ($i=20^{\circ}$, $50^{\circ}$ and
$80^\circ$) are summarised in Fig.~\ref{fig:example-MA-DW-all}.

At an intermediate inclination angle ($i=50^{\circ}$), 
the \ion{He}{i}~$\lambda$10830 profile in Fig.~\ref{fig:example-MA-DW-all}
shows the absorption features in both red and blue wings. The relatively
narrow blueshifted absorption component is caused by the disc wind
(outflow), and the absorption in the red wing is caused by the magnetospheric
accretion funnel (inflow). The narrow blueshifted absorption component
is commonly found in the observed \ion{He}{i}~$\lambda$10830 (e.g.~$\sim30$~per~cent
of the sample in \citealt{Edwards:2006}). Similarly, the redshifted
absorption component is also commonly (47~per~cent) seen in the
observed \ion{He}{i}~$\lambda10830$ profiles of \citet{Edwards:2006}. 

The figure shows that \ion{He}{i}~$\lambda$5876 profiles are much
weaker than \ion{He}{i}~$\lambda10830$ profiles, with this particular
set of model parameters.  The line ratio of \ion{He}{i}~$\lambda$5876
and \ion{He}{i}~$\lambda10830$ (using their peak fluxes) is about
$\sim0.07$ for the line profiles computed at the intermediate
inclination angle $i=50^\circ$. This is very similar to the line
ratios found in the low temperature gas models of \citet{Kwan:2011}
(e.g. see their Figure~9). The relative strength of
\ion{He}{i}~$\lambda$5876 is expected to be greater in a model with
higher temperature gas and higher density than those used here.  Our
model profiles show the redshifted absorption components for all the
inclination angles even although the absorption is very weak for a low
inclination angle model.  On the other hand, the observations of
\ion{He}{i}~$\lambda$5876 profiles, from 31 CTTSs obtained by
\citet*{Beristain:2001}, show that the redshifted absorption is rather
rare (3 out 31). The occurrence of the redshifted absorption in the
model may be decreased when a smaller mass-accretion rate was used
(e.g.~$10^{-8}\,\MsunPerYear$). Both our models and the observations
do not show the blueshifted absorption in \ion{He}{i}~$\lambda$5876.

The H$\alpha$ profile shows broad wings (extending up to $\pm400\,\mathrm{\kmps}$)
which are mainly caused by the broadening mechanism explained in Section~\ref{sub:obs-prof}.
It also shows a relatively narrow blueshifted absorption component,
caused by the disc wind. However, unlike \ion{He}{i}~$\lambda$10830,
the redshifted absorption below the continuum is not seen in this
H$\alpha$ profile. Based on the morphology of the profile, the profile
is classified as `Type III-B' according the classification scheme
of \citet{reipurth:1996}. This type of line profile is found in about
33 per~cent of the observed H$\alpha$ (\citealt{reipurth:1996}). 

Interestingly, the wind signature (the blueshifted absorption component)
is absent in the model profiles of Pa$\beta$ and Br$\gamma$. In
both lines, only the redshifted absorption component, caused by the
accreting gas in the magnetosphere, is present. This type of line
profile is often referred to as the inverse P-Cygni (IPC) profiles,
and is well reproduced by the magnetospheric accretion model of \citet{hartmann:1994}
(see also \citealt*{Muzerolle:1998}; \citealt{muzerolle:2001}; \citealt{symington:2005};
\citealt*{kurosawa:2005}; \citealt{kurosawa:2006}). This is consistent
with the near infrared spectroscopic observations of 50 T Tauri stars
by \citet{folha:2001} who found an almost complete absence of blueshifted
absorption components and a relatively high frequency of IPC profiles
in Pa$\beta$ and Br$\gamma$. 
Similarly, the wind signature is absent in Pa$\gamma$ profiles
shown in Fig.~\ref{fig:example-MA-DW-all}. This agrees with the
observation of \citet{Edwards:2006} who presented the atlas of
Pa$\gamma$ profiles from 38 CTTSs, which also shows little or no
signature of wind absorption components. The morphology of the
Pa$\gamma$ model profiles shown here overall agrees with the ranges of the
profile shapes found in the observation of \citet{Edwards:2006}. 

As mentioned earlier, the sensitivity of \ion{He}{i}~$\lambda10830$
to the innermost winds of CTTSs and its usefulness for proving the
physical conditions of the winds have been well demonstrated in the
past (e.g.~\citealt{Edwards:2006}; \citealt{Kwan:2007}; \citealt{Edwards:2009};
\citealt{Kwan:2011})\@. 
To test whether the magnetosphere--disc wind hybrid model
can account for the types of \ion{He}{i}~$\lambda10830$ profiles
seen in the observations, we have run a small set of profile models
for a various combinations of $\dot{M}_{\mathrm{dw}}$,
$T_{\mathrm{dw}}$ and $i$. The rest of the parameters are kept same as
in our canonical model shown in Fig.~\ref{fig:example-MA-DW-all}. 
We then examined if there is a resemblance between any of our model
profiles to a set of \ion{He}{i}~$\lambda10830$ line observations in
\citet{Edwards:2006}. In the following, we present an example case in
which the model profile morphology resembles that of an observation.

\begin{figure}

\includegraphics[clip,width=0.48\textwidth]{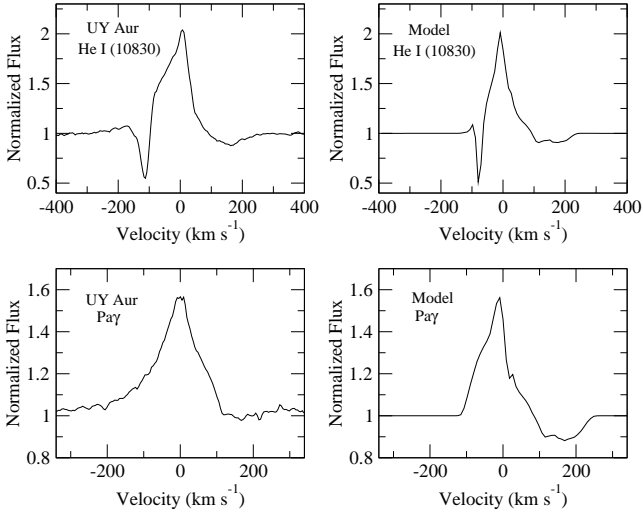}

\caption{
      Example model calculations with a combination of
      axisymmetric magnetosphere and disc wind. 
      The normalized model profiles (right panels) are compared with the
      observed line profiles (left panels) of UY~Aur obtained by
      \citet{Edwards:2006}.  The upper and lower panels show
      \ion{He}{i}\,$\lambda$10830 and Pa$\gamma$ profiles,
      respectively. The observed profiles are obtained simultaneously
      (see \citealt{Edwards:2006} for detail).  The inclination angle
      used for computing the model profiles is
      $i=48^{\circ}$. The model parameters adopted here are:
      $T_{\mathrm{max}}=$7,500~K, $T_{\mathrm{dw}}=$9,000~K,
      $L_{\mathrm{X}}=2\times10^{29}\,\mathrm{erg s^{-1}}$, 
      $\dot{M}_{\mathrm{a}}=10^{-7}\,\MsunPerYear$, and 
      $\dot{M}_{\mathrm{dw}}=2.5\times10^{-9}\,\MsunPerYear$. N.B.~the
      model line profiles shown here are not strict fits to the observations, but
      are simple comparisons of the profile morphology.
} 

\label{fig:example-MA-DW}

\end{figure}

Fig.~\ref{fig:example-MA-DW} shows a simple comparison of a
\ion{He}{i}~$\lambda10830$ line profile from our magnetosphere--disc
wind hybrid model with the observed \ion{He}{i}~$\lambda10830$ profile
of UY~Aur (CTTS; M0) obtained by \citealt{Edwards:2006}. The model profile in the figure
is computed with the following parameters: $T_{\mathrm{dw}}=$~9,000~K,
$\dot{M}_{\mathrm{dw}}=2.5\times10^{-9}\,\MsunPerYear$ and $i=48^\circ$. The 
mass-accretion rate for UY~Aur adopted here
($\log\dot{M}_{\mathrm{a}}=-7.0$) is similar to the
observationally estimated value of $\log\dot{M}_{\mathrm{a}}=-7.2$
(\citealt{Edwards:2006}), but the adopted wind mass-loss rate is
slightly smaller than the observed value of
$\log\dot{M}_{\mathrm{w}}=-8.2$ (\citealt{Edwards:2006}). The model
reproduces the basic features of the observed
\ion{He}{i}~$\lambda10830$ line profile remarkably well. We find good
matches of the model and the observation in: (1)~the location and the
narrowness of the blueshifted absorption line (caused by the disc
wind), (2)~the line strength, and (3)~the depth and the width of the
redshifted absorption (caused by the magnetospheric accretion). This
clearly demonstrates the plausibility of our model (magnetosphere + disc
wind) in explaining the type of line profile exhibited by UY~Aur.  Our
model agrees with the earlier finding by \citet{Kwan:2007} and
\citet{Edwards:2009} who suggest that the relatively narrow
blueshifted absorption seen in the observed \ion{He}{i}~$\lambda10830$
line of UY~Aur is likely caused by a disc wind.

In addition to \ion{He}{i}~$\lambda10830$, we have also computed
Pa$\gamma$ profile for UY~Aur since the observation presented by
\citet{Edwards:2006} also includes Pa$\gamma$ (which is obtained
simultaneously with \ion{He}{i}~$\lambda10830$). The result is shown
also in Fig.~\ref{fig:example-MA-DW}. Both model and the observation
shows no sign of blueshifted wind absorption.  A small amount of
redshifted absorption is seen in both model and the observation.  The
overall agreement between the model and observed Pa$\gamma$ profiles
is very good.  An important difference between the model and
observation is in the strength of the blue wing emission. The
observation shows a very notable amount of emission in the wing
($V<-100\,\kmps$). However, the model do not show this emission. This
may suggest that the emission from the wind in UY~Aur is much stronger
than that of our model. To match the extended blue wing, we may need
to adjust the wind temperature to a higher value. A further discussion
on the blue wing emission will be given in Section~\ref{sub:issues}.

Again, our main objective here is to test whether the magnetosphere--disc
wind hybrid model can roughly account for the types of
\ion{He}{i}~$\lambda10830$ profiles seen in the observations, but is
not to derive a set of physical parameters that is required to
reproduce the observed line profile as this requires more careful
analysis. In other words, the model profile shown here is not a strict
fit to the observation. However, given the similarity of our model
profiles with those of UY~Aur, we should be able to derive the physical
parameters of the accretion flow and wind around the object by
performing the fine grid parameter search and performing the
$\chi^{2}$ fitting to not only \ion{He}{i}~$\lambda10830$ line but
also to other hydrogen and helium lines, simultaneously.  This is
beyond the scope of this paper, but shall be left for a future
investigation.

\subsection{Stellar wind model}

\label{sub:stellar-wind-model}

\subsubsection{Configuration}

\label{sub:stellar-wind-config}

As shown in the observations of \ion{He}{i}~$\lambda10830$ by \citet{Edwards:2006},
about 40~per~cent of the CTTSs in their sample exhibit P-Cygni profiles.
The observations indicate a possible presence of rapidly expanding
stellar winds, in some of CTTSs. The stellar wind threaded by the
open magnetic field in the polar region is a possible mechanism for
angular momentum loss (e.g.~\citealt{Hartmann:1982b}; \citealt{Mestel:1984};
\citealt{hartmann:1989}; \citealt{Tout:1992}; \citealt{Paatz:1996};
\citealt{matt:2005}; \citealt{Matt:2008a}; \citealt{Matt:2008b};
\citealt{Matt:2010}). Recent work by \citet{Cranmer:2008} and \citet{Cranmer:2009}
have shown a possible mechanism of heating and acceleration of the stellar
wind, in the context of CTTSs, driven by accretion.

We replace the disc wind, described in Section~\ref{sub:discwind-config},
with the stellar wind, which arises only in the polar directions as
shown in Fig.~\ref{fig:config-MA-SW}. Our stellar wind model consists
of narrow cones with the half-opening angle $\theta_{\mathrm{sw}}$
in the polar directions ($\pm z$ direction as shown in the figure).
The wind is assumed to be propagating only in the radial direction,
and its velocity is described by the classical beta-velocity law (cf.~\citealt{castor:1979})
i.e. 
\begin{equation}
  v_{r}\left(r\right)=v_{0}+\left(v_{\infty}-v_{0}\right)\left(1-\frac{R_{0}}{r}\right)^{\beta}\,,
  \label{eq:beta-velocity-law}
\end{equation}
where $v_{\infty}$ and $v_{0}$ are the terminal velocity and the
velocity of the wind at the base ($r=R_{0}$). Assuming the mass-loss
rate by the wind is $\dot{M}_{\mathrm{sw}}$ and using the mass-flux
conservation in the flows in cones, the density $\rho_{\mathrm{sw}}$
of the wind can be written as: 
\begin{equation}
  \rho_{\mathrm{sw}}\left(r\right)=\frac{\dot{M}_{\mathrm{sw}}}{4\pi
    r^{2}v_{r}\left(r\right)\left(1-\cos\theta_{\mathrm{sw}}\right)}\,,
  \label{eq:sw-density}
\end{equation}
where $\theta_{\mathrm{sw}}>0$, and $\rho_{\mathrm{sw}}$ becomes
that of a spherical wind when $\theta_{\mathrm{sw}}=90^{\circ}$.
The temperature of the stellar wind ($T_{\mathrm{sw}}$) here is also
assumed isothermal, as in the case for the disc wind model (Section~\ref{sub:discwind-config}). 

When the wind is launched from the stellar surface (i.e.~$R_{0}=R_{*}$),
the half-opening angle $\theta_{\mathrm{sw}}$ is restricted to be
less than $\sim35^{\circ}$ to avoid an overlapping of the wind with
the accretion funnels (cf. Fig.~\ref{fig:config-MA-SW}). On the
other hand, if we assume that the wind is launched at a slightly larger,
e.g. at the outer radius of the magnetosphere $R_{0}=R_{\mathrm{mo}}$,
the opening angle of the wind is not restricted by the geometry of
the accretion funnels. 
In the models presented in this paper, we adopt $R_{0}=R_{\mathrm{mo}}
(= 3.0\,R_{*})$ and $\theta_{\mathrm{sw}}=50^{\circ}$.

\subsubsection{Example profiles}

\label{sub:ma-sw-example}

\begin{figure}

\begin{center}

\includegraphics[clip,width=0.4\textwidth]{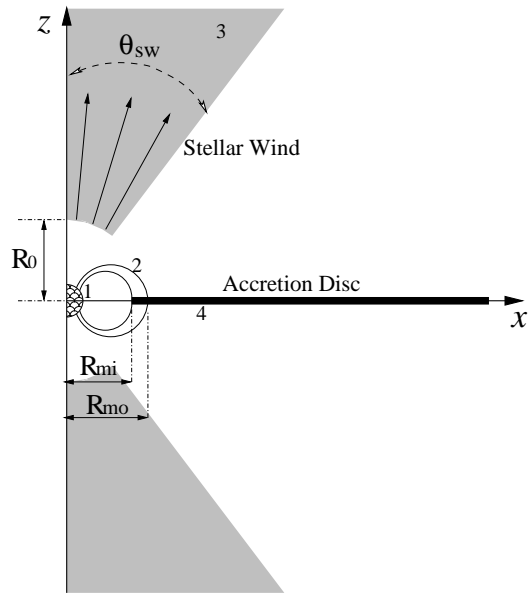}

\end{center}

\caption{Same as in Fig.~\ref{fig:config-MA-DW}, but for the (bipolar)
  stellar wind and magnetospheric accretion hybrid model. The system
  consists of four components: (1)~the continuum source located at the
  origin of the cartesian coordinates $\left(x,y,z\right)$ -- the
  $y$-axis is into the paper, (2)~the magnetospheric accretion flow,
  (3)~the stellar wind, and (4)~the optically thick but geometrically
  thin accretion disc. 
  The wind is launched from a sphere with radius
  $R_{0}$, but is restricted within the cones with the half opening
  angle $\theta_{\mathrm{sw}}$. We adopt $R_{0}=R_{\mathrm{mo}} (=3.0~R_{*})$ and
  $\theta_{\mathrm{sw}}=50^{\circ}$ for the models presented in this
  paper.  
The density distribution is symmetric around the $z$-axis. The figure
is not to scale.}  

\label{fig:config-MA-SW}

\end{figure}

\begin{figure}

\includegraphics[clip,width=0.48\textwidth]{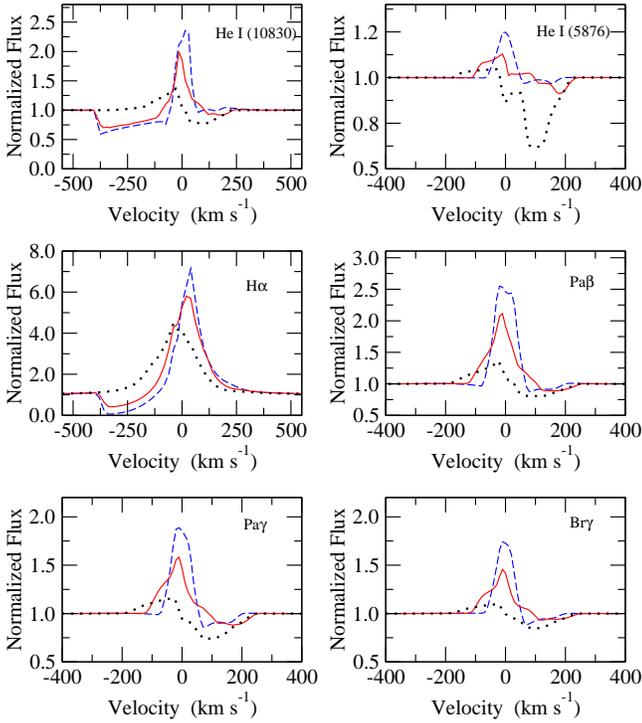}

\caption{An example model calculation with a combination of
  axisymmetric magnetosphere and a stellar wind. The normalized model
  profile are shown as a function of velocity. The model parameters
  adopted here are: $T_{\mathrm{max}}=$~7,500~K,
  $T_{\mathrm{sw}}=$~8,000~K,
  $\dot{M}_{\mathrm{a}}=10^{-7}\,\MsunPerYear$,
  $\dot{M}_{\mathrm{sw}}=10^{-8}\,\MsunPerYear$, and
  $L_{\mathrm{X}}=2\times10^{29}\,\mathrm{erg\,s^{-1}}$.  The profiles
  are computed at inclination angles $i=20^{\circ}$ (dashed),
  $50^{\circ}$ (solid) and $80^{\circ}$ (dotted).
    N.B. the lower limits of the vertical axis for \ion{He}{i}~(10830)
    and H$\alpha$ are set to 0, but those for \ion{He}{i}~(5876),
    Pa$\beta$, Pa$\gamma$ and Br$\gamma$ are set to 0.5.
} 

\label{fig:example-MA-SW-all}

\end{figure}

Fig.~\ref{fig:example-MA-SW-all} shows the example helium and hydrogen
profiles computed for the standard axisymmetric magnetospheric accretion
funnels ($R_{\mathrm{mi}}=2.2\, R_{*}$, $R_{\mathrm{mo}}=3.0\, R_{*}$,
$\dot{M}_{\mathrm{a}}=10^{-7}\,\MsunPerYear$ and $T_{\mathrm{max}}=7,500\,\Kelvin$)
and the stellar wind with the following parameters: $T_{\mathrm{sw}}=8,000\,\Kelvin$,
$\dot{M}_{\mathrm{sw}}=10^{-8}\,\MsunPerYear$, $\theta_{\mathrm{sw}}=50^{\circ}$,
$v_{0}=10\,\kmps$, $v_{\infty}=400\,\kmps$,  $R_{0}=3.0\, R_{*}$, and
$\beta=0.5$ (cf.~Section~\ref{sub:stellar-wind-config} and Fig.~\ref{fig:config-MA-SW}).
The parameters for the central star are as described in Section~\ref{sub:cont-photosphere}.
The effective temperature of the hotspot, as described in Section~\ref{sub:cont-hotspot},
is approximately 6,400~K. The X-ray luminosity of the chromosphere
(cf.~Section~\ref{sub:cont-xray}) is assumed to be $L_{\mathrm{X}}=2\times10^{29}\,\mathrm{erg\, s^{-1}}$.
The profiles shown here are computed for the system with the inclination
angle of $i=53^{\circ}$. Note that the wind half opening angle
$\theta_{\mathrm{sw}}$ here is much wider ($\sim50^{\circ}$) than the
one shown in Fig.~\ref{fig:config-MA-SW} since the base of the stellar wind in this particular
model is set to be at the outer radius of the magnetosphere
($R_{\mathrm{0}}=R_{\mathrm{mo}}$) instead of the stellar surface
($R_{\mathrm{0}}=R_{\mathrm{*}}$). 
The stellar and disk occultations are taken into account in our profile calculations.

As in the disc wind cases (Section~\ref{sub:ma-dw-example}), we
have adopted a relatively low stellar wind temperature (8,000~$\Kelvin$)
here to avoid unrealistically strong emission in the hydrogen lines,
produced at the base of the wind. Again, for the same reason, the
temperature of the magnetosphere cannot be much higher than $T_{\mathrm{max}}=7,500\,\Kelvin$
(for the system with $\dot{M}_{\mathrm{a}}=10^{-7}\,\MsunPerYear$).
As done for the disc wind case, we have added the X-ray flux (as described
in Section~ \ref{sub:cont-xray}) in the stellar radiation as an
additional source of photoionization. This helps to populate the excited
states of \ion{He}{i} and to produce a reasonable line strength in
\ion{He}{i}~$\lambda$10830.  The resulting profiles are shown in
Fig.~\ref{fig:example-MA-SW-all}. 

The \ion{He}{i}~$\lambda$10830 profiles
(Fig.~\ref{fig:example-MA-SW-all}), computed at the low and
intermediate inclination angles ($i=20^{\circ}$ and $50^{\circ}$),
show the absorption features in both red and blue wings.  However,
unlike the disc wind case, the blueshifted absorption is much wider
and it extends almost to the terminal velocity of the wind,
$v_{\infty}=400\,\kmps$.  This is the classical P-Cygni profile which
is caused by the expanding stellar wind. As mentioned earlier, about
40~per~cent of the CTTSs in the observations of \citet{Edwards:2006}
show the P-Cygni profiles in \ion{He}{i}~$\lambda10830$. The model
profile of the \ion{He}{i}~$\lambda10830$ also shows the absorption in
the red wing, which is caused by the magnetospheric accretion funnel
(inflow). Again, the redshifted absorption component is also commonly
(47~per~cent) found in the observed \ion{He}{i}~$\lambda10830$
profiles of \citet{Edwards:2006}.

The model line profiles of \ion{He}{i}~$\lambda$5876
(Fig.~\ref{fig:example-MA-SW-all}) are very similar to those computed
with the disc wind (Fig.~\ref{fig:example-MA-DW-all}). Again, there is
no clear sign of a blueshifted wind absorption in the models.  The
similarity of the profiles here with those of the disk wind model
(Fig.~\ref{fig:example-MA-DW-all}) and the fact that the physical
conditions (the density, temperature and velocity) of the
magnetospheres used here are exactly the same as in the models with
the disc wind (Fig.~\ref{fig:example-MA-DW-all}) suggests that, at
least with this particular set of model parameters, the emission in
\ion{He}{i}~$\lambda$5876 mainly occurs in the magnetosphere.  As in
the disc wind case, \ion{He}{i}~$\lambda$5876 profiles are much weaker
than \ion{He}{i}~$\lambda10830$ profiles.  The line ratio of
\ion{He}{i}~$\lambda$5876 and \ion{He}{i}~$\lambda10830$ (using their
peak fluxes) is about $\sim0.07$ for the line profiles computed at the
intermediate inclination angle $i=50^\circ$.  As mentioned earlier,
the redshifted absorption in \ion{He}{i}~$\lambda$5876 is rather rare
(about 10 per~cent) in the observation of \citet{Beristain:2001} (see
also \citealt{Alencar:2000}).

The H$\alpha$ profile here also shows a broad wing in red, extending
up to $+400\,\mathrm{\kmps}$. The blue wing is strongly affected by
the presence of the wind absorption component 
for the models computed at $i=50^{\circ}$ and $80^{\circ}$.
Similar to \ion{He}{i}~$\lambda$10830,
the blueshifted absorption in the H$\alpha$ profile is wide and it
extends almost to $v_{\infty}=400\,\kmps$. This profile can be also
classified as a P-Cygni profile. Unlike \ion{He}{i}~$\lambda$10830,
the redshifted absorption does not go below the continuum level in
this H$\alpha$ profile. Based on the morphology of the profile, it
is classified as `Type IV-B' according the classification scheme of
\citet{reipurth:1996}. This type of line profile is relatively rare
($\sim$7~per~cent) in the observed H$\alpha$ profiles of \citet{reipurth:1996}. 

As we found in the disc wind case (Section~\ref{sub:discwind-model}),
the wind signature (the blueshifted absorption component) is absent
in the model profiles of Pa$\beta$ and Br$\gamma$. Only the redshifted
absorption component cased by the accreting gas in the magnetosphere
is present (the IPC profiles). This is consistent with the near infrared
spectroscopic observations of 50 T Tauri stars by \citet{folha:2001}.
The model profiles of Pa$\beta$ and Br$\gamma$ shown here are very similar to
those of the disc wind model (in Fig.~\ref{fig:example-MA-DW-all})
because these lines are not affected by the presence of the wind,
and the inclinations ($i$) used in both models are same. 
The wind signature is also absent in Pa$\gamma$ profiles shown in
Fig.~\ref{fig:example-MA-SW-all}. As mentioned earlier, this is
consistent with the observation of \citet{Edwards:2006}. Again, the
morphology of the Pa$\gamma$ model profiles shown here overall agrees
with the ranges of the profile shapes found in the observation of
\citet{Edwards:2006}.

As in the disc wind case, we now test whether the
magnetosphere--stellar wind hybrid model can roughly account for the
types of \ion{He}{i}~$\lambda10830$ profiles seen in the
observations. We have run a small set of profile models for a
various combinations of $\dot{M}_{\mathrm{sw}}$, $T_{\mathrm{sw}}$,
$i$, and the turbulence velocity in the magnetosphere and the wind. 
Note that no turbulence velocity is included in the previous models.
The rest of the parameters are kept same as in our canonical
model shown in Fig.~\ref{fig:example-MA-SW-all}.  We then examined if
there is a resemblance between any of our model profiles to a set of
\ion{He}{i}~$\lambda10830$ line observations in \citet{Edwards:2006}.

\begin{figure}

\includegraphics[clip,width=0.48\textwidth]{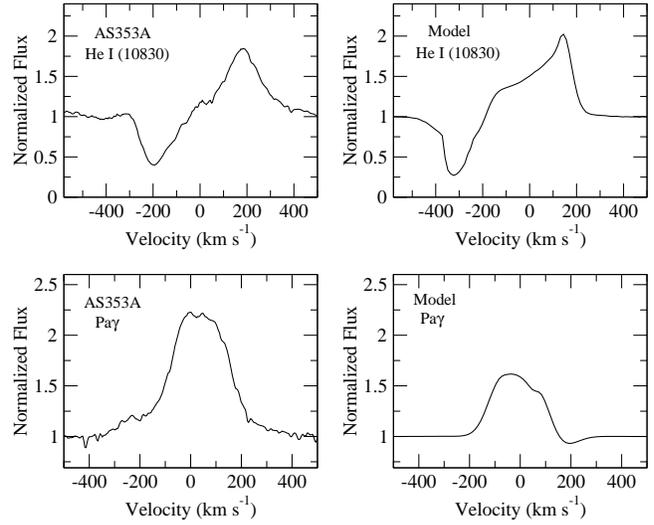}

\caption{
      Example model calculations with a combination of
      axisymmetric magnetosphere and a stellar wind. 
      The normalized model profiles (right panels) are compared with the
      observed line profiles (left panels) of  AS~353~A  obtained by
      \citet{Edwards:2006}.  The upper and lower panels show
      \ion{He}{i}\,$\lambda$10830 and Pa$\gamma$ profiles,
      respectively. The observed profiles are obtained simultaneously
      (see \citealt{Edwards:2006} for detail).  The inclination angle
      used for computing the model profiles is
      $i=10^{\circ}$. The model parameters adopted here are:
      $T_{\mathrm{max}}=$7,500~K, $T_{\mathrm{sw}}=$8,000~K,
      $L_{\mathrm{X}}=1\times10^{29}\,\mathrm{erg s^{-1}}$, 
      $\dot{M}_{\mathrm{a}}=10^{-7}\,\MsunPerYear$, and 
      $\dot{M}_{\mathrm{sw}}=2\times10^{-9}\,\MsunPerYear$. N.B.~the
      model line profiles shown here are not strict fits to the observations, but
      are simple comparisons of the profile morphology.
}
\label{fig:example-MA-SW}

\end{figure}

Fig.~\ref{fig:example-MA-SW} shows a simple comparison of a \ion{He}{i}~$\lambda10830$
line profile from our magnetosphere--stellar wind hybrid model with
the observed \ion{He}{i}~$\lambda10830$ profile of AS~353~A (CTTS;
K5) from \citet{Edwards:2006}. The model profile in the figure is computed with the
following parameters: $T_{\mathrm{sw}}=$~8,000~K,
$\dot{M}_{\mathrm{\mathrm{sw}}}=2.5\times10^{-8}\,\MsunPerYear$,
$L_{\mathrm{X}}=1\times10^{29}\,\mathrm{erg\,s^{-1}}$, 
$i=10^\circ$,  and the turbulence velocity of 80~$\kmps$ in both the
magnetosphere and the wind. 
The rest of the parameters are same as those used in
Fig.~\ref{fig:example-MA-SW-all}.  Note that the mass-accretion rate
for AS~353~A adopted here ($\log\dot{M}_{\mathrm{a}}=-7.0$) is much smaller than the observationally
estimated value of $\log\dot{M}_{\mathrm{a}}=-5.4$
(\citealt{Edwards:2006}), but the adopted wind mass-loss rate is very
smaller to the observed value of $\log\dot{M}_{\mathrm{w}}=-7.6$
(\citealt{Edwards:2006}). 

By following \citet{Kwan:2007}, to match the location of the
emission peak in red ($\sim200\,\kmps$), a relatively high turbulence
velocity (80~$\kmps$ in both the magnetosphere and the wind)
is added to the model. 
  Note that \citet{Kwan:2007} applied a high turbulence
  velocity in the wind only, but not in the magnetosphere --- their
  model does not include a magnetosphere at all.  As we
  demonstrate later in Section~\ref{sub:issues}
  (Fig.~\ref{fig:emiss-loc}), the line `emission' mainly arises from
  the magnetosphere in the models presented in this paper;
  therefore, it is not possible to match the peak of the observed
  profile without invoking a significant (80~$\kmps$) turbulence
  velocity also in the magnetospheric accretion funnels.  This
  essentially arbitrary line-broadening (over-and-above thermal and
  pressure broadening) may have a physical basis in the
  magnetically-channelled nature of the inflows and outflows, and
  further radiative transfer models will be necessary to investigate
  whether it would be possible to determine whether the different
  turbulent broadening may apply in both the wind and the accretion
  streams. If one could find a solution in which the wind emission
  dominates, the high turbulence velocity in the magnetosphere assumed
  here might not be necessary.

The figure shows that the
magnetosphere--stellar wind hybrid model reproduces the basic features
of the observed \ion{He}{i}~$\lambda10830$ line profile relatively
well.  The model profile shows a 
clear P-Cygni profile, and the overall agreement of the model and the
observational line profile morphology is very well. Again, this
demonstrates the plausibility of our model (magnetosphere + stellar wind)
in explaining the type of line profile exhibited by AS~353~A.
Our model agrees with the earlier finding by \citet{Kwan:2007} and
\citet{Edwards:2009} who suggest that the relatively wide blueshifted
absorption (the P-Cygni profile) seen in the observed
\ion{He}{i}~$\lambda10830$ line of AS~353~A is likely caused by a
stellar wind. 

We have also computed Pa$\gamma$ profile for AS~353~A, and compared it
with the observed profile from \citet{Edwards:2006}. The data is
acquired simultaneously with the \ion{He}{i}~$\lambda10830$ line. The
profiles are also shown in Fig.~\ref{fig:example-MA-SW}. No clear sign
of wind absorption is seen in both model and observed Pa$\gamma$
profiles. 
  Note that the line width of the model Pa$\gamma$ profile in
  Fig.~\ref{fig:example-MA-SW} is much larger than that of the
  Pa$\gamma$ profile in Fig.~\ref{fig:example-MA-SW-all} (for
  $i=20^\circ$ case) because of the high turbulence velocity
  (80~$\kmps$) in the magnetosphere used in the model for AS~353~A.
The model shows a weak redshifted absorption, but the
observation does not. 
The line strength of the model is noticeably weaker than the observed profile.
Another important difference between the model and observation is
in the amount of the emission in the wings of the
profiles. The observation shows significant emissions in the wing
extending up to $V\sim \pm 400\,\kmps$.  On the other hand, our model
do not show strong wing emissions.  This difference was also found in
the comparison of Pa$\gamma$ profiles in UY~Aur
(Fig.~\ref{fig:example-MA-DW}).  
To produce the extended wing emission 
and a stronger line centre flux in our model,
a higher wind temperature may be needed.  A further
discussion on the extended wing emissions will be given in
Section~\ref{sub:issues}.

Once again, our main objective here is to demonstrate whether the
magnetosphere--stellar wind hybrid model can roughly account for the
types of \ion{He}{i}~$\lambda10830$ profiles seen in the observations,
but is not to derive a set of physical parameters that is required to
reproduce the observed line profile as this requires more careful
analysis. For example, the mass-accretion rate used here is much
smaller than the observed value and it should not be considered as a
derived model parameter.  However, the resemblance of the model
profiles to those of AS~353~A suggests a possibility for deriving the
physical parameters of the accretion flow and wind around the object
by performing the fine grid parameter search and performing the
$\chi^{2}$ fitting to a set of (preferably) simultaneously observed
multiple hydrogen and helium lines (including
\ion{He}{i}~$\lambda10830$).

\section{Discussions}
\label{sec:discussions}

\subsection{Some unresolved issues}
\label{sub:issues}

\begin{figure}

\begin{center}
\includegraphics[clip,width=0.48\textwidth]{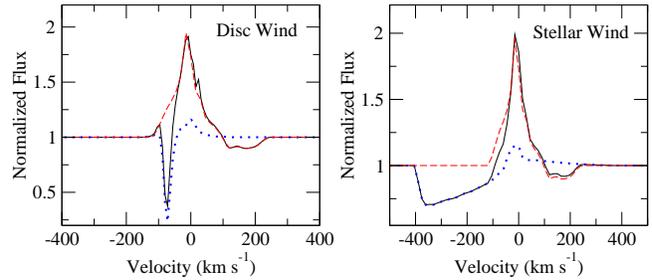}
\par\end{center}

\caption{Comparisons of model \ion{He}{i}~$\lambda10830$ line profiles
  computed with different flow elements in the disc wind +
  magnetosphere configuration  (left panel) and in the stellar
  wind + magnetosphere  configuration (right panel). The profiles are
  computed with (1)~the magnetosphere only (dashed), (2)~the
  disc/stellar wind only (dotted), and (3)~the combination of the
  magnetosphere and disc/stellar wind (solid). The
  physical parameters used here are the same as those in
  Figs.~\ref{fig:example-MA-DW-all} and \ref{fig:example-MA-SW-all}
  for the models with the disc wind and that with the stellar wind,
  respectively. The inclination angle $i=50^\circ$ is used for all the
  models. 
} 

\label{fig:emiss-loc}

\end{figure}

While the blue and redshifted absorption components in the model line
profiles (e.g. in Figs.~\ref{fig:example-MA-DW-all} and
\ref{fig:example-MA-SW-all}) could provide us the physical conditions
and locations of line `absorption', the locations of line `emission'
are not so straightforwardly understood from the model profiles. To
demonstrate which gas flow component (magnetosphere or wind)
contributes to the emission part of the profiles, we compare the
\ion{He}{i}~$\lambda10830$ line profiles computed using (1)~the
magnetosphere only, (2)~the disc/stellar wind only, and (3)~the
combination of the magnetosphere and disc/stellar wind.  The results
are placed in Fig.~\ref{fig:emiss-loc}.  The physical parameters used
here are the same as those in Figs.~\ref{fig:example-MA-DW-all} and
\ref{fig:example-MA-SW-all} for the models with the disc wind and that
with the stellar wind, respectively. As one can see from the figure,
with these particular sets of model parameters, mainly the
magnetospheres contribute to the emission. The wind emission is
relatively low, and the winds mainly contribute to blueshifted
absorptions. Although not shown here, the magnetosphere is also a main
emission contributor in other lines shown in
Figs.~\ref{fig:example-MA-DW-all} and \ref{fig:example-MA-SW-all}.
Since we have not yet surveyed a large parameter space for the wind
and magnetosphere in our models, we cannot generalise the statement
above. The relative importance of the wind and magnetospheric emission
and the determination of the emission location require further
investigation which we defer to a future study.

There are at least two important differences between the model
profiles, computed with the magnetosphere as a main line emitter, and
observed profiles. Firstly, the line widths of Pa$\beta$ and
Br$\gamma$ in our models (e.g.~Figs.~\ref{fig:example-MA-DW-all} and
\ref{fig:example-MA-SW-all}) are much narrower than those from the
observation by \citet{folha:2001}.  The full width half-maxima (FWHM)
of Pa$\beta$ and Br$\gamma$ are about 200~$\kmps$ in
\citet{folha:2001}, but those in our model are $\sim100$ or
less. Since the maximum infall velocity in our model is essentially
limited to the escape velocity from the star ($\sim300\,\kmps$ with
$R_{*}=2\,R_{\odot}$ and $M_{*}=0.5\,M_{\odot}$), it is hard to
achieve the FWHM values found in the observation without an additional
form of line broadening (e.g.~turbulence
broadening). \citet{kurosawa:2006} also discuss a similar problem. The
emission from a stronger wind may also help to make the lines broader,
but the stronger wind would also cause a blueshifted absorption.

Secondly, the extent of the blue wing emission in the model profiles
are much smaller than those found in observations. In the comparison
of the model profiles with the observation of UY~Aur
(Fig.~\ref{fig:example-MA-DW}), we found the blue wings of the
observed \ion{He}{i}~$\lambda10830$ and Pa$\gamma$ extend up to $\sim
300\,\kmps$ while our models show the blue wings only up to $\sim
130\,\kmps$. Similarly for AS~353~A (see
Fig.~\ref{fig:example-MA-SW}), the blue wing of the observed
Pa$\gamma$ profile extends up to $\sim 400\,\kmps$ while that of the
model extends only up to $~200\,\kmps$, even with a large turbulence
velocity ($80\,\kmps$) included.  As mentioned earlier, in the models
considered here, the line emission mainly arises in the
magnetosphere. Hence, with this set up, the models have difficulties
in producing the blue wing emission extending to such high velocity.
This discrepancy may suggest that the emission from a fast wind (a
disc or stellar wind) may be necessary to reproduce the extended blue
wing emission seen in the emission.  Whether one can reproduce the
strong blue wing emission by a wind without causing a strong
blueshifted absorption and a (too) strong central emission component
is still unknown. This issue will be investigated further in a future
study.

As we have briefly mentioned in Section~\ref{sub:ma-sw-example},
H$\alpha$ lines computed with the stellar wind + magnetospheric
accretion (see Fig.~\ref{fig:example-MA-SW-all}) exhibit strong
P-Cygni like blueshifted absorptions; however, this type of profile
(Type IV-B in the classification by \citealt{reipurth:1996}) is rare
in the observations. In our model configuration (with a stellar wind),
it is difficult not to produce a P-Cygni like absorption when we
require a P-Cygni like absorption in \ion{He}{i}~$\lambda10830$. In
other words, whenever a P-Cygni like absorption is present in
\ion{He}{i}~$\lambda10830$, it also appears in H$\alpha$.  One
possible way to avoid this problem is to use a lower wind temperature
which reduces the ionization level of hydrogen in the wind, but not
that of helium since the helium ionization level is mainly controlled
by the X-ray flux. Another possibility is to increase the temperature
of the magnetosphere which would increase the emission in the wings of
H$\alpha$ and would suppress the strong blue absorption features. This
point shall also be investigated further in a future study.

\subsection{X-ray optical depths}
\label{sub:x-ray-tau}

When computing the statistical equilibrium of the atomic population
levels, we have neglected the attenuation of X-ray emission by
implicitly assuming the radiation is optically thin. Here we examine
whether this assumption is reasonable, by computing the X-ray optical
depth (continuum) using the ionization structure found in the disc
wind plus magnetospheric accretion model shown in
Fig.~\ref{fig:example-MA-DW-all} (Section~\ref{sub:ma-dw-example}) and
also the stellar wind plus magnetospheric accretion model shown in
Fig.~\ref{fig:example-MA-SW-all}
(Section~\ref{sub:ma-sw-example}). Note that the optical depths
computed from these models are not self-consistent since the inclusion
of the X-ray attenuation should, in principle, alter the ionization
structures itself, especially for \ion{He}{i}. Hence, the following
optical depth values are merely rough estimates.

For the photon energy $\epsilon=0.1$~keV and along a line of sight to
an observer at $i=50^\circ$, the optical depth across the
magnetospheric accretion stream is $\tau_{\mathrm{X}}\approx 10^3$,
and that across the wind is $\tau_{\mathrm{X}}\approx 20$ and $\approx
40$ for the disc wind and stellar winds, respectively. On the other
hand, for $\epsilon=1.0$~keV, $\tau_{\mathrm{X}}\approx 1$ for the
magnetosphere, and $\tau_{\mathrm{X}}\approx 0.01$ and $\approx 0.02$
for the disc and stellar winds respectively.  The optical depths are
computed with the models with the mass-accretion
$\dot{M}_{\mathrm{a}}=10^{-7}\,\MsunPerYear$ and the wind mass-loss
rate $\dot{M}_{\mathrm{w}}=10^{-8}\,\MsunPerYear$, which are about 10
times higher than those of a typical CTTS. If more typical values are
used (i.e.~$\dot{M}_{\mathrm{a}}=10^{-8}\,\MsunPerYear$ and
$\dot{M}_{\mathrm{w}}=10^{-9}\,\MsunPerYear$), the optical depths
would be about 10 times lower than the values shown above.  In our
model, the main source of the continuum opacity around the photon
energy 0.1~keV is the bound-free opacity of \ion{He}{i}.

The high optical depths at 0.1~keV indicates that the most of the
X-ray at this energy could not penetrate the high density
magnetospheric accretion streams. On the other hand, the optical depth
across the magnetosphere at 1.0~keV is marginal
($\tau_{\mathrm{X}}\sim 1.0$), and may penetrate the accretion
streams, after its intensity being reduced by a factor of
$e^{\tau_{\mathrm{X}}} \sim 2.7$. This suggests that our \ion{He}{i}
line strength may be overestimated for a given amount of X-ray
luminosity.  In reality, the magnetospheric accretion may not be
completely axisymmetric, and the accretion would occur only in two or
more funnel streams as shown by 3D MHD simulations
(e.g.~\citealt{romanova:2003}). This indicates a presence of large
gaps between the funnel flows; hence, a significant fraction of the
X-ray produced near the stellar surface could still reach to the
stellar or disc wind by escaping though the gaps, and ionize
\ion{He}{i} in the wind.  We plan to implement a proper treatment of
X-ray attenuation effect in a future model.

\section{Summary and conclusions}

\label{sec:conclusions}

The radiative transfer code {\sc torus} (\citealt{harries:2000};
\citealt{kurosawa:2004}; \citealt{symington:2005}: \citealt{kurosawa:2006};
\citealt{kurosawa:2008}) is extended to include helium atoms (19
levels for \ion{He}{i}, 10 levels for \ion{He}{ii} and the continuum
level for \ion{He}{iii}). Only hydrogen atoms (up to 20 levels) are
included in the previous versions of the code. The addition
of helium atoms in our code is motivated by the recent observations
of \citet{Edwards:2003}, \citet{dupree:2005} and \citet{Edwards:2006}
who demonstrated a robustness of optically thick \ion{He}{i}~$\lambda10830$
as a diagnostic tool for probing the innermost wind of CTTSs. 

We have tested our new implementation of the statistical equilibrium
routine which simultaneously solves H and He level populations (Section~\ref{sub:test-1D})
by comparing the model with the non-LTE line-blanketed atmosphere
code {\sc cmfgen} (\citealt{hillier:1998}) in which the level populations
are solved in a spherical geometry (1D). For the case of spherically
expanding hot stellar wind (AV~83),  we have found a good agreement
between {\sc torus} and {\sc cmfgen} in computing the number fractions of He
ions (Fig.~\ref{fig:ion-frac}), opacities (Fig.~\ref{fig:opacity}),
and observed line profiles (Fig.~\ref{fig:test1d-profile}). 
The code has been also tested in a 2D (axisymmetric) geometry (Section~\ref{sub:test-2D}).
The emission line profile models of \citet{muzerolle:2001}, who used
the axisymmetric magnetospheric accretion model of \citet{hartmann:1994},
have been well reproduced with our model (Fig.~\ref{fig:compare-mch}).

We have also demonstrated the capabilities of our newly implemented
code in the line formation problems of CTTSs
(Section~\ref{sec:apps-to-ctts}).  We have considered two different
combinations of the inflow and outflow geometries: (1)~the
magnetospheric accretion and the disc wind
(Section~\ref{sub:discwind-model}), and (2)~the magnetospheric
accretion and the bipolar stellar wind
(Section~\ref{sub:stellar-wind-model}). In both cases, our models are
able to produce line profiles similar to the observed ones
(e.g.~\citealt{edwards:1994}; \citealt{reipurth:1996};
\citealt{folha:2001}; \citealt{Edwards:2006}), in the morphology and
the line strengths (Figs.~\ref{fig:example-MA-DW-all} and
\ref{fig:example-MA-SW-all}).  In particular, the comparison of the
model \ion{He}{i}~$\lambda10830$ profiles with with the observations
of \citet{Edwards:2006} (Figs.~\ref{fig:example-MA-DW} and
\ref{fig:example-MA-SW}) suggests that the relatively deep and wide
blueshifted absorption component found in some CTTSs is most likely
caused by the bipolar stellar wind while the narrower blueshifted
absorption is caused by the disc wind (cf.~\citealt{kurosawa:2006};
\citealt{Kwan:2007}).

During our initial investigation of the simultaneous modelling of H
and He emission lines, we found (in both the disc wind and stellar
wind cases) the temperature of the wind must be relatively low
e.g.~$T\sim$10,000~K (for $\dot{M}_{\mathrm{a}}=10^{-7}\,\MsunPerYear$
and it could be slightly higher for a lower mass-accretion rate), and
it is likely below 20,000~K. The line strengths of hydrogen lines
(e.g.~H$\alpha$) will be unrealistically strong if the temperature is
much higher. On the other hand, at this relatively low wind
temperature ($T\sim$10,000~K), the collisional rates of \ion{He}{i}
are too low, and the normal continuum sources (the photosphere with
$T_{\mathrm{eff}}\sim$4,000~K and the hot spot with
$T_{\mathrm{hs}}\sim$6,400~K or even with
$T_{\mathrm{hs}}\sim$8,000~K) do not provide enough radiation to
ionize/excite \ion{He}{i} significantly.  To populate the excited
states of \ion{He}{i} and hence to produce \ion{He}{i}~$\lambda10830$
with a realistic line strength (c.f.~Figs.~\ref{fig:example-MA-DW} and
\ref{fig:example-MA-SW}), one requires a high energy radiation source
such as X-ray radiation.  Interestingly, a recent work by
\citet{Kwan:2011} has shown that UV photoionization is also a probable
excitation mechanism for generating \ion{He}{i}~$\lambda10830$
opacities. From the observations of TW~Hya and RU Lup,
\citet{Johns-Krull:2007} also reached a similar conclusion. Both our
model and the work by \citet{Kwan:2011} suggest a need for the extra
photoionizing radiation sources. The relative importance of the UV and
X-ray luminosity is not known at this moment.

The following is the list of a several key questions which are yet
to be answered: Do the stellar wind and disc wind coexist? In which
physical condition, one dominates the other? What are the possible
ranges of temperature and density in the winds? Must the winds of
CTTS be clumpy (inhomogeneous), as suggested by \citet{Kwan:2011}?
What is the relative importance of the UV and X-ray? Where exactly
is the location of X-ray emission? We intent to investigate and provide
answers to some or most of the questions mentioned above in our subsequent
papers in the future. 

Finally, we also plan to apply our radiative transfer model to the
results of 2D and 3D MHD simulations of the (innermost) conical wind
and magnetospheric accretion of CTTSs by \citet{Romanova:2009}.
A direct comparison of the MHD simulations with observations is difficult,
but it can be mediated by modelling emission line profiles which could
constrain some basic physical parameters of the flow such as the geometry,
temperature and mass accretion and outflow rates. In turn, these parameters
can provide constraints on the amount of angular momentum transfer
to and out of the stars. By using the different time slices of the
MHD simulations, we will be also able to address the issues of the
time-variability in the line and continuum by comparing the models
with observations (e.g.~\citealt{kurosawa:2008}).

\section*{Acknowledgements} 

We thank an anonymous referee who provided us valuable comments and
suggestions which helped improving the manuscript.
We also thank D.~John Hillier for making the O star wind model and the
atomic data available online, and for the helpful comments on the
manuscript. We also thank Susan Edwards for providing us the observed
data of \ion{He}{i}~$\lambda10830$ and Pa$\gamma$ line profiles, and
for the helpful comments on the manuscript.  
The research of RK and MMR is supported by NSF grant AST-0807129 and
NASA grant NNX10AF63G.


\end{document}